\documentclass[pra,aps,nofootinbib,notitlepage,superscriptaddress,twocolumn]{revtex4-1}

\usepackage{amsmath}
\usepackage{amssymb}
\usepackage{amsfonts}
\usepackage{graphicx}

\usepackage{txfonts}
\usepackage{upgreek}

\newcommand{\supp}[1]{Supplementary Note #1}
\newcommand{\bs}[1]{\boldsymbol{#1}}

\newcommand{\ms}[1]{\mathit{#1}}

\newcommand{\vectheta}{\bs{\uptheta}}
\newcommand{\matSigma}{\varSigma}
\newcommand{\matPi}{\varPi}
\newcommand{\matM}{\mathcal{M}}
\newcommand{\matA}{\ms{A}}
\newcommand{\matB}{\ms{B}}
\newcommand{\matC}{\ms{C}}
\newcommand{\matD}{\ms{D}}
\newcommand{\matE}{\ms{E}}
\newcommand{\matF}{\ms{F}}
\newcommand{\matG}{\ms{G}}
\newcommand{\matK}{\ms{K}}
\newcommand{\matR}{\ms{R}}
\newcommand{\matS}{\ms{S}}
\newcommand{\matP}{\ms{P}}
\newcommand{\matO}{\ms{O}}
\newcommand{\matU}{\ms{U}}
\newcommand{\matT}{\ms{T}}
\newcommand{\matW}{\ms{W}}
\newcommand{\matV}{\ms{V}}
\newcommand{\matY}{\ms{Y}}
\newcommand{\matID}{\mathit{1}}
\newcommand{\matzero}{\mathit{0}}
\newcommand{\matGamma}{\varGamma}
\newcommand{\matOmega}{\varOmega}
\newcommand{\matomega}{\omega}
\newcommand{\matxi}{\varXi}

\usepackage[colorlinks=true,linkcolor=blue,citecolor=blue,urlcolor=blue]{hyperref}

\begin{document}

\title{Multiparameter squeezing for optimal quantum enhancements in sensor networks}

\author{Manuel Gessner}
\affiliation{Laboratoire Kastler Brossel, ENS-PSL Universit\'{e}, CNRS, Sorbonne Universit\'{e}, Coll\`{e}ge de France, 24 Rue Lhomond, 75005 Paris, France}
\author{Augusto Smerzi}
\affiliation{QSTAR, CNR-INO and LENS, Largo Enrico Fermi 2, I-50125 Firenze, Italy}
\author{Luca Pezz\`e}
\affiliation{QSTAR, CNR-INO and LENS, Largo Enrico Fermi 2, I-50125 Firenze, Italy}
\date{\today}

\maketitle

\textbf{
Squeezing currently represents the leading strategy for quantum enhanced precision measurements of a single parameter in a variety of continuous- and discrete-variable settings and technological applications. However, many important physical problems including imaging and field sensing require the simultaneous measurement of multiple unknown parameters. The development of multiparameter quantum metrology is yet hindered by the intrinsic difficulty in finding saturable sensitivity bounds and feasible estimation strategies. Here, we derive the general operational concept of multiparameter squeezing, identifying metrologically useful states and optimal estimation strategies. When applied to spin- or continuous-variable systems, our results generalize widely-used spin- or quadrature-squeezing parameters. Multiparameter squeezing provides a practical and versatile concept that paves the way to the development of quantum-enhanced estimation of multiple phases, gradients, and fields, and for the efficient characterization of multimode quantum states in atomic and optical sensor networks.
}
\section*{Introduction}
Squeezing of quantum observables is a central strategy to improve measurement sensitivities beyond classical limits and has thus become a key concept in quantum metrology, leading to major theoretical and experimental advancements in the field~\cite{RMP, MaPHYSREP2011, TothJPA, Caves, KitagawaUeda, Wineland}. Furthermore, squeezing is a convenient approach to witness genuine quantum properties such as entanglement~\cite{SMPRL01, Sorensen} or nonclassicality~\cite{WallsBOOK}, only requiring knowledge of first and second moments of suitable linear observables that can be obtained experimentally with high efficiency. The concept of squeezing is most useful for the important class of Gaussian states that is routinely generated in atomic and photonic experiments~\cite{BraunsteinVanLoock,Ferraro,Wang,Weedbrook,RMP}.

While well understood in the framework of single-parameter estimation~\cite{RMP, MaPHYSREP2011, TothJPA, Caves, Wineland}, 
the existing notion of squeezing is insufficient to characterize the sensitivity of multiparameter estimation. 
Indeed, the simultaneous estimation of several parameters 
can be more efficient than the optimal estimation of each parameter 
separately~\cite{HumphreysPRL2013, GessnerPRL2018, ProctorPRL2018, GePRL2018}. 
This interesting prediction is under 
intensive investigation~\cite{CiampiniSC2016, RagyPRA2016, LiuJPA2016, GagatsosPRA2016,  NicholsPRA2018, Guo2019} 
and can revolutionize many technological applications such as quantum imaging \cite{KolobovBOOK}, microscopy and astronomy \cite{TsangPRX2016,TsangPRL2016,LupoPRL2016,RehacekPRA2017}, 
sensor networks \cite{ProctorPRL2018,GePRL2018,Guo2019} and atomic clocks~\cite{Lukin}, by enhancing the estimation sensitivity of inhomogeneous intensity distributions, vector fields, and gradients \cite{WildermuthAPL2006,Koschorreck,UrizarPRA2013,AltenburgPRA2017,ApellanizPRA2018}. However, the current framework of multiparameter quantum metrology has developed based on the notion of the quantum Fisher information matrix~\cite{HelstromBOOK}: a figure of merit that is not straightforward to extract experimentally and is also generally hard to determine theoretically. 
Furthermore, the sensitivity limit defined by the inverse of the quantum Fisher information matrix, 
namely, the multiparameter quantum Cram\'{e}r-Rao bound~\cite{HelstromBOOK}, 
is, in general, not saturable~\cite{MatsumotoJPA2002,PezzePRL2017}. 
Alternative approaches based on the Holevo bound are in principle asymptotically saturable but
require, in general, complex measurements on multiple copies of the state~\cite{Holevo,Yamagata,Yang,Albarelli,AlbarelliReview}.

In this work, we introduce the general notion of metrological multiparameter squeezing for continuous and discrete variables.
This concept follows directly from a specific operational approach to multiparameter estimation based on mean values and variances of the measured observables.
Metrological multiparameter squeezing thus provides an accessible and saturable lower bound to the quantum Fisher matrix that is tight for 
the broad and experimentally-relevant class of Gaussian states. We further use matrix order inequalities to analytically optimize the measurement observables as a function of accessible observables. Our framework is neither limited to specific systems nor to a particular class of observables and provides an efficient characterization of useful quantum resources for 
multiparameter estimation for any given set of commuting observables that are simultaneously measured. For linear spin observables our method gives rise to the spin-squeezing matrix as a natural generalization of the spin-squeezing coefficient introduced by Wineland \textit{et al.}~\cite{Wineland} to multiparameter settings. The spin-squeezing matrix reveals the role of nonlocal squeezing, i.e., squeezing in a nonlocal superposition of modes for simultaneous estimations of multiple parameters
which can enhance the sensitivity of specific linear combinations of parameters. We further identify optimal strategies for displacement sensing in continuous variables, where nonlocal squeezing over $M$ modes can reduce the estimation error up to a factor $\sqrt{M}$. To address the properties of non-Gaussian states, we demonstrate that our approach can yield a multiparameter sensitivity as large as the classical Fisher matrix (and even the quantum Fisher matrix, whenever the multiparameter quantum Cram\'{e}r-Rao bound is saturable).

\section*{Results}

\textbf{Multiparameter method of moments.}
In multiparameter quantum metrology~\cite{HelstromBOOK} the goal is to estimate a family of unknown parameters $\vectheta=(\theta_1,\dots,\theta_M)^T$. The parameters are imprinted onto $\hat{\rho}$ by a unitary evolution $\hat{U}(\vectheta)=\exp(-i\hat{\mathbf{H}}\cdot\vectheta)=\exp(-i\sum_{k=1}^M\hat{H}_k\theta_k)$, where $\hat{\mathbf{H}}=(\hat{H}_1,\dots,\hat{H}_M)^T$ is a vector of Hamiltonians that do not necessarily commute with each other. After the phase imprinting, a measurement is performed and the experiment is repeated $\mu$ times with the same output state $\hat{\rho}(\vectheta)=\hat{U}(\vectheta)\hat{\rho}\hat{U}(\vectheta)^{\dagger}$. The parameters $\theta_k$ are inferred from a set of estimators $\theta_{\mathrm{est},k}$ with $k=1,\dots,M$, which are functions of the measurement results. The multiparameter uncertainty is quantified by the $M\times M$ covariance matrix $\matSigma$ with elements $\matSigma_{kl}=\mathrm{Cov}(\theta_{\mathrm{est},k},\theta_{\mathrm{est},l})$. The operational meaning of $\matSigma$ is that, for an arbitrary $M$-dimensional real vector of coefficients $\mathbf{n} = (n_1, \dots, n_M)^T$, the quantity $\mathbf{n}^T\matSigma\mathbf{n} = \Delta^2 (n_1 \theta_{{\rm est},1} + \cdots + n_M \theta_{{\rm est},M})$ yields the variance of the corresponding linear combination of estimators.

We introduce here an estimation protocol based on a multiparameter method of moments. The parameters $\vectheta$ are estimated from the average values of a set of $K$ measurement observables $\hat{\mathbf{X}}=(\hat{X}_1,\dots,\hat{X}_K)^T$. We consider a commuting set $\hat{\mathbf{X}}$ to ensure simultaneous measurability in a single shot, but our framework does not formally require this assumption. In the central limit we obtain the covariance matrix (see Methods for details) 
\begin{align}\label{eq:gaussianfishermatrix}
\matSigma=(\mu\matM[\hat{\rho}(\vectheta),\hat{\mathbf{H}},\hat{\mathbf{X}}])^{-1}.
\end{align}
The moment matrix
\begin{align}\label{eq:momentmatrix}
\matM[\hat{\rho}(\vectheta),\hat{\mathbf{H}},\hat{\mathbf{X}}]=\matC[\hat{\rho}(\vectheta),\hat{\mathbf{H}},\hat{\mathbf{X}}]^T \, \matGamma[\hat{\rho}(\vectheta),\hat{\mathbf{X}}]^{-1} \,
\matC[\hat{\rho}(\vectheta),\hat{\mathbf{H}},\hat{\mathbf{X}}],
\end{align}
depends on the covariance matrix $(\matGamma[\hat{\rho}(\vectheta),\hat{\mathbf{X}}])_{kl}=\langle\hat{X}_k\hat{X}_l\rangle_{\hat{\rho}(\vectheta)}-\langle\hat{X}_k\rangle_{\hat{\rho}(\vectheta)}\langle\hat{X}_l\rangle_{\hat{\rho}(\vectheta)}$ and the commutator matrix $(\matC[\hat{\rho}(\vectheta),\hat{\mathbf{H}},\hat{\mathbf{X}}])_{kl}=-i\langle [\hat{X}_k,\hat{H}_l]\rangle_{\hat{\rho}(\vectheta)}$. 
Equation (\ref{eq:momentmatrix}) provides a lower bound to the classical and quantum Fisher information matrix, i.e., 
\begin{align} \label{eq:MF}
\matM[\hat{\rho}(\vectheta),\hat{\mathbf{H}},\hat{\mathbf{X}}]\leq\matF[\hat{\rho}(\vectheta),\hat{\mathbf{X}}]\leq \matF_Q[\hat{\rho}(\vectheta),\hat{\mathbf{H}}],
\end{align}
expressing, e.g., that $\matF-\matM$ is a positive semidefinite matrix~\cite{Stein}. The classical Fisher matrix $\matF[\hat{\rho}(\vectheta),\hat{\mathbf{X}}]$ determines the multiparameter sensitivity limit~\cite{HelstromBOOK,KayBOOK} attainable by a measurement of the observables $\hat{\mathbf{X}}$ and consists of elements 
$(\matF[\hat{\rho}(\vectheta),\hat{\mathbf{X}}])_{kl}=\sum_{\mathbf{x}}p(\mathbf{x}|\vectheta)\left(\frac{\partial}{\partial \theta_k}\log p(\mathbf{x}|\vectheta)\right)\left(\frac{\partial}{\partial \theta_l}\log p(\mathbf{x}|\vectheta)\right)$,
where $p(\mathbf{x}|\vectheta)=\mathrm{Tr}\{\hat{\Pi}_{\mathbf{x}}\hat{\rho}(\vectheta)\}$ is the probability to obtain the result $\mathbf{x}=(x_1,\dots,x_K)^T$ and the $\hat{\bs{\Pi}}=\{\hat{\Pi}_{\mathbf{x}}\}_{\mathbf{x}}$ denote the projectors onto the common eigenstates of the $\hat{X}_k$. 
For any fixed basis, defined by the projectors $\hat{\bs{\Pi}}$, the bound (\ref{eq:MF}) can be saturated by an optimal choice of the measurement observables $\hat{\mathbf{X}}$ (e.g., by measuring directly the projectors $\hat{\mathbf{X}}=\hat{\bs{\Pi}}$), leading to
\begin{align}\label{eq:boundMF}
\max_{\hat{\mathbf{X}}\in\mathrm{span}(\hat{\bs{\Pi}})}\matM[\hat{\rho}(\vectheta),\hat{\mathbf{H}},\hat{\mathbf{X}}]=\matF[\hat{\rho}(\vectheta),\hat{\mathbf{X}}].
\end{align}
The short-hand notation $\hat{\mathbf{X}}\in\mathrm{span}(\hat{\bs{\Pi}})$ expresses that each of the $\hat{X}_k$ is a linear combination of the elements of $\hat{\bs{\Pi}}$. Moreover, the bound $\matF[\hat{\rho}(\vectheta),\hat{\mathbf{X}}]\leq \matF_Q[\hat{\rho}(\vectheta),\hat{\mathbf{H}}]$ holds for all $\hat{\mathbf{X}}$, where 
$(\matF_Q[\hat{\rho},\hat{\mathbf{H}}])_{kl}=\mathrm{Tr}\{\hat{\rho}(\hat{L}_k\hat{L}_l+\hat{L}_l\hat{L}_k)/2\}$ is the quantum Fisher information \cite{HelstromBOOK,BraunsteinPRL1994} 
and $-i[\hat{H}_k,\hat{\rho}]=(\hat{L}_k\hat{\rho}+\hat{\rho}\hat{L}_k)/2$ defines the symmetric logarithmic derivative operators. 
The Fisher information matrix equals the quantum Fisher  matrix only under certain conditions \cite{MatsumotoJPA2002,PezzePRL2017}.
Equations~(\ref{eq:MF}) and~(\ref{eq:boundMF}) and their saturation conditions are derived in \supp{2}. The bounds~(\ref{eq:MF}) show that the moment matrix~(\ref{eq:momentmatrix}) approximates the state's multiparameter sensitivity by means of first and second moments of the chosen measurement observables $\hat{\mathbf{X}}$. For linear observables $\hat{\mathbf{X}}$ (e.g., collective spins or quadratures), this can be interpreted as a Gaussian approximation of the (quantum) Fisher matrix, but through the measurement of nonlinear observables the method is also able to efficiently characterize non-Gaussian states.

In the following, we present an analytical method for identifying the optimal choice of $\hat{\mathbf{X}}$. We consider here the case of a predefined family of accessible operators $\hat{\mathbf{A}}=(\hat{A}_1,\dots,\hat{A}_L)^T$ with $L\geq M,K$, that may be chosen as the experimentally available observables. The optimization will be realized under the constraint that only linear combinations of the operators $\hat{\mathbf{A}}$ can be measured. We thus assume that $\hat{\mathbf{X}}$, as well as the Hamiltonians $\hat{\mathbf{H}}$, can be expressed as linear combinations $\hat{H}_k=\sum_{i=1}^Lr_{k,i}\hat{A}_i$ and $\hat{X}_k=\sum_{i=1}^Ls_{k,i}\hat{A}_i$. The real-valued coefficients $r_{k,i}$ and $s_{k,i}$ define the $M\times L$ and $K\times L$ transformation matrices $\matR$ and $\matS$, respectively. We may write 
\begin{align}\label{eq:HXRS}
\hat{\mathbf{H}}=\matR\hat{\mathbf{A}},\quad\text{and}\quad\hat{\mathbf{X}}=\matS\hat{\mathbf{A}},
\end{align}
and henceforth we assume $\matR\matR^T=\matID_M$ and $\matS\matS^T=\matID_K$. We first optimize the choice of the matrix $\matS$, i.e., the measurement observables, for any fixed phase encoding transformation specified by the matrix $\matR$.
The optimization of the moment matrix~(\ref{eq:momentmatrix}) is given by
\begin{align}\label{eq:maxM}
\matM_{\mathrm{opt}}[\hat{\rho},\hat{\mathbf{H}},\hat{\mathbf{A}}]:=\max_{\hat{\mathbf{X}}\in\mathrm{span}(\hat{\mathbf{A}})}\matM[\hat{\rho},\hat{\mathbf{H}},\hat{\mathbf{X}}]=\matR \tilde{{\matM}}[\hat{\rho},\hat{\mathbf{A}}] \, \matR^T,
\end{align}
where 
\begin{align}\label{eq:momentmatrixA}
\tilde{{\matM}}[\hat{\rho},\hat{\mathbf{A}}]=\tilde{\matC}[\hat{\rho},\hat{\mathbf{A}}]^T\matGamma[\hat{\rho},\hat{\mathbf{A}}]^{-1}\tilde{\matC}[\hat{\rho},\hat{\mathbf{A}}]
\end{align}
is the $L\times L$ moment matrix of operators $\hat{\mathbf{A}}$, which is defined on the basis of the covariance matrix $(\matGamma[\hat{\rho},\hat{\mathbf{A}}])_{kl}=\frac{1}{2}\langle\hat{A}_k\hat{A}_l+\hat{A}_l\hat{A}_k\rangle_{\hat{\rho}}-\langle\hat{A}_k\rangle_{\hat{\rho}}\langle\hat{A}_l\rangle_{\hat{\rho}}$ and the commutator matrix $(\tilde{\matC}[\hat{\rho},\hat{\mathbf{A}}])_{kl}=-i\langle[\hat{A}_k,\hat{A}_l]\rangle_{\hat{\rho}}$. The result~(\ref{eq:maxM}) is proven in \supp{2} and follows from the matrix inequality
\begin{align}\label{eq:upperbound}
\matM[\hat{\rho},\hat{\mathbf{H}},\hat{\mathbf{X}}]\leq \matR\tilde{{\matM}}[\hat{\rho},\hat{\mathbf{A}}]\matR^T,
\end{align}
which holds for arbitrary $\hat{\mathbf{X}}$. Saturation in~(\ref{eq:upperbound}) is achieved by the observables defined in Eq.~(\ref{eq:HXRS}) if and only if there exists a real-valued $K\times M$ matrix $\matG$ such that
\begin{align}\label{eq:optimalX}
\matG\matS=\matR\tilde{\matC}[\hat{\rho},\hat{\mathbf{A}}]^T\matGamma[\hat{\rho},\hat{\mathbf{A}}]^{-1}.
\end{align}
This result generalizes the analytical optimization discussed in Ref.~\cite{GessnerPRL2019} to the multiparameter case. Moreover, the choice of parameter-encoding Hamiltonians, i.e., $\matR$ can be optimized by considering the spectrum of $\tilde{{\matM}}[\hat{\rho},\hat{\mathbf{A}}]$ (see Methods). In practice, the optimal moment matrix~(\ref{eq:maxM}) can only be achieved by a direct measurement if the elements of an optimal $\hat{\mathbf{X}}$, defined by~(\ref{eq:optimalX}), can be measured simultaneously.

{\bf Squeezing matrix.} We define the squeezing matrix by comparing the moment-based sensitivity $\matSigma$ of Eq.~(\ref{eq:gaussianfishermatrix}) to the multiparameter shot-noise limit $\matSigma_{\mathrm{SN}}$, i.e., the sensitivity limit of classical measurement strategies. While this approach can be applied to arbitrary multiparameter estimation scenarios, in the following we focus mostly on the experimentally relevant cases of distributed sensor networks or multimode interferometers~\cite{ProctorPRL2018,GePRL2018,GessnerPRL2018}: The parameters are encoded in $M$ different modes by local Hamiltonians satisfying $[\hat{H}_k,\hat{H}_l] = 0$ for all $k,l$ and we measure one observable in each mode $(K=M)$. In these cases, the shot-noise limit $\matSigma_{\mathrm{SN}}=(\mu \matF_{\mathrm{SN}}[\hat{\mathbf{H}}])^{-1}$ can be explicitly determined from the quantum Cram\'er-Rao bound (see Methods). For evolutions generated by $\hat{\mathbf{H}}$ and measurement observables $\hat{\mathbf{X}}$ we define the squeezing matrix as
\begin{align}\label{eq:squeezingmatrix}
\matxi^{2}[\hat{\rho},\hat{\mathbf{H}},\hat{\mathbf{X}}]:=\matF_{\mathrm{SN}}[\hat{\mathbf{H}}]^{\frac{1}{2}}\matM[\hat{\rho},\hat{\mathbf{H}},\hat{\mathbf{X}}]^{-1}\matF_{\mathrm{SN}}[\hat{\mathbf{H}}]^{\frac{1}{2}}.
\end{align}
By expressing Eq.~(\ref{eq:gaussianfishermatrix}) as $\matSigma=\matSigma_{\mathrm{SN}}^{\frac{1}{2}}\matxi^{2}[\hat{\rho},\hat{\mathbf{H}},\hat{\mathbf{X}}]\matSigma_{\mathrm{SN}}^{\frac{1}{2}}$, we observe that the squeezing matrix $\matxi^{2}[\hat{\rho},\hat{\mathbf{H}},\hat{\mathbf{X}}]$ directly quantifies the quantum gain in a saturable, moment-based multiparameter estimation protocol. Any quantum state with the property
\begin{align}\label{eq:SQZcond}
\matxi^{2}[\hat{\rho},\hat{\mathbf{H}},\hat{\mathbf{X}}]\geq \matID_M,
\end{align}
can only yield multiparameter shot-noise sensitivity or worse, i.e., $\matSigma\geq \matSigma_{\mathrm{SN}}$. Inserting Eq.~(\ref{eq:maxM}) into~(\ref{eq:squeezingmatrix}), we obtain the optimized squeezing matrix:
\begin{align} \label{eq:xiopt}
\matxi_{\mathrm{opt}}^{2}[\hat{\rho},\hat{\mathbf{H}},\hat{\mathbf{A}}]&:=\min_{\hat{\mathbf{X}}\in\mathrm{span}(\hat{\mathbf{A}})}\matxi^{2}[\hat{\rho},\hat{\mathbf{H}},\hat{\mathbf{X}}]\notag\\
&=\matF_{\mathrm{SN}}[\hat{\mathbf{H}}]^{\frac{1}{2}}\matR\tilde{{\matM}}[\hat{\rho},\hat{\mathbf{A}}]^{-1}\matR^T\matF_{\mathrm{SN}}[\hat{\mathbf{H}}]^{\frac{1}{2}}.
\end{align}

A violation of the matrix inequality~(\ref{eq:SQZcond}) 
signals multiparameter squeezing (with respect to the phase-imprinting Hamiltonians $\hat{\mathbf{H}}$ and the measurement observables $\hat{\mathbf{X}}$): it
implies that there exists at least one vector $\mathbf{n}\in\mathbb{R}^M$ for which 
$\mathbf{n}^T\matSigma\mathbf{n}<\mathbf{n}^T\matSigma_{\mathrm{SN}}\mathbf{n}$ holds. In this case, sub-shot-noise sensitivity is achieved for the estimation of $\mathbf{n}^T\vectheta$, which describes a particular linear combination of the parameters. The number $0\leq r_{\mathrm{SN}}\leq M$ of negative eigenvalues of the matrix $\matSigma - \matSigma_{\mathrm{SN}}$ defines the shot-noise rank \cite{GessnerPRL2018} that is achieved by the multiparameter method of moments. Equivalently, $r_{\rm SN}$ corresponds to the number of eigenvalues of $\matxi^2$ that are smaller than one. 
When $r_{\mathrm{SN}}=M$, the stronger condition $\matxi^{2}[\hat{\rho},\hat{\mathbf{H}},\hat{\mathbf{X}}]< \matID_M$ for 
full multiparameter squeezing is satisfied. In this case $\matSigma<\matSigma_{\mathrm{SN}}$ holds, and sub-shot noise sensitivity is achieved for the estimation of arbitrary $\mathbf{n}^T\vectheta$.

The observation of multiparameter squeezing implies that the state is nonclassical (see Methods). To increase the quantum enhancements, it is thus beneficial to reduce the squeezing matrix as much as possible by using strongly nonclassical states. 

\begin{figure}[t!]
\centering
\includegraphics[width=.49\textwidth]{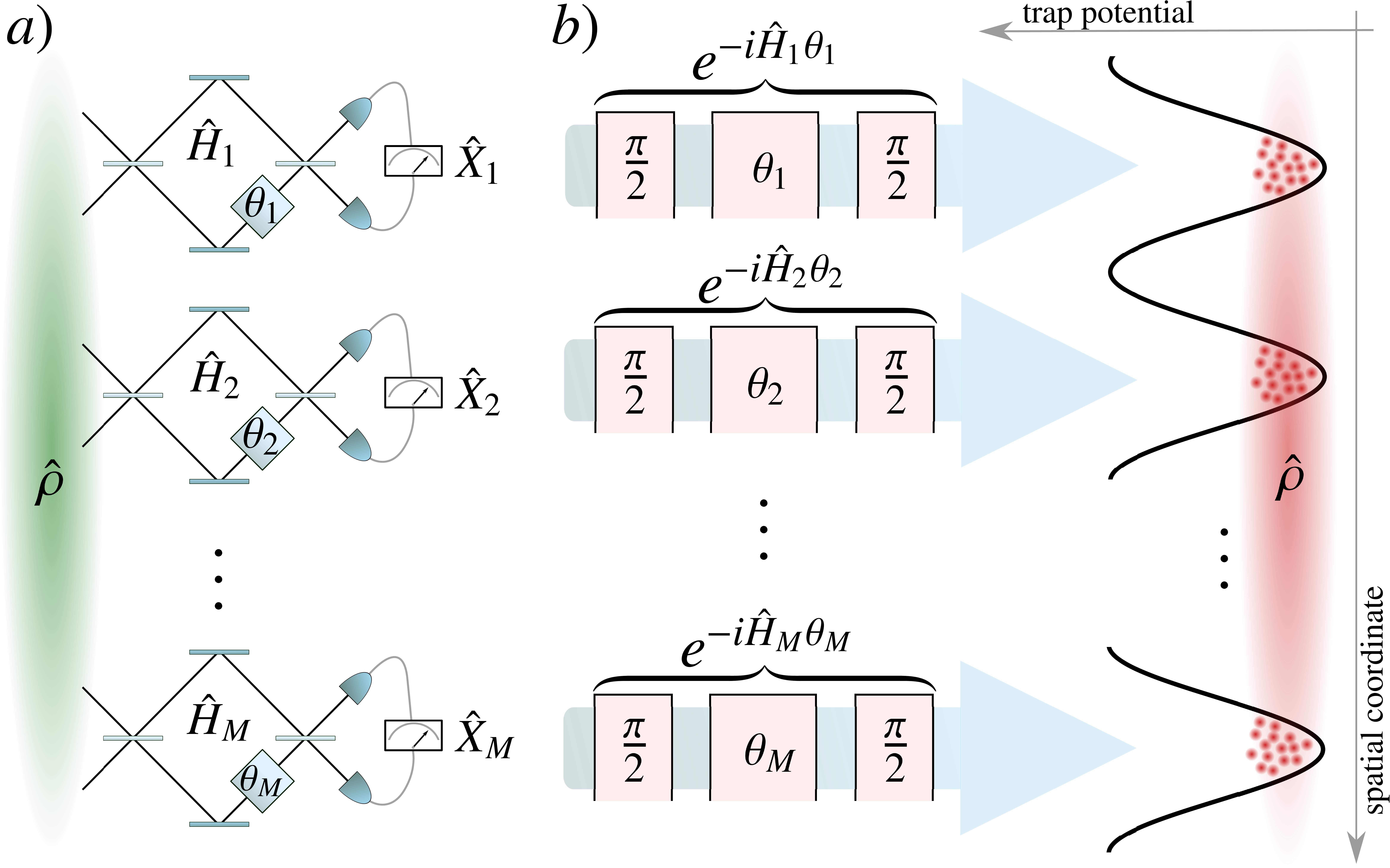}
\caption{Quantum-enhanced parallel interferometers. In each mode $k=1,\dots,M$ of a set of Mach-Zehnder (a) or Ramsey interferometers (b), a single parameter $\theta_k$ is imprinted by a local Hamiltonian $\hat{H}_k$, and a local observable $\hat{X}_k$ is measured. The multiparameter sensitivity is quantified by the moment matrix~(\ref{eq:momentmatrix}). The multiparameter quantum gain is captured by the squeezing matrix~(\ref{eq:squeezingmatrix}), which contains both local (single-parameter) enhancements and nonlocal (multiparameter) squeezing. The sensitivity can be optimized analytically using Eq.~(\ref{eq:maxM}) and the maximum is achieved when Eq.~(\ref{eq:optimalX}) is fulfilled for a set of commuting observables $\hat{X}_1,\dots,\hat{X}_M$.}
\label{fig:MMZ}
\end{figure}

{\bf Multiparameter discrete-variable (spin) squeezing.} 
Discrete-variable multiparameter estimation provides the theoretical framework to model a series of $M$ local Ramsey or Mach-Zehnder 
interferometers that operate in parallel, each with a fixed number of particles $N_k$, with $k=1,...,M$; see Fig.~\ref{fig:MMZ}. 
Here, each mode is modeled by a collective spin of length $N_k/2$, for $k=1,\dots,M$, summing up to a total number of $N=\sum_{k=1}^MN_k$ spin-$1/2$ particles. 
The multimode interferometer is described by a family of local parameter-encoding Hamiltonians $\hat{\mathbf{H}}=\hat{\mathbf{J}}_{\mathbf{r}}=(\hat{J}_{\mathbf{r}_1,1},\dots,\hat{J}_{\mathbf{r}_M,M})^T$, where $\mathbf{r}=(\mathbf{r}_1,\dots,\mathbf{r}_M)$, $\hat{J}_{\mathbf{r}_k,k}=\mathbf{r}_k^T\hat{\mathbf{J}}_{\perp,k}$, $\hat{\mathbf{J}}_{\perp,k}=(\hat{J}_{x,k},\hat{J}_{y,k})^T$, and $\hat{J}_{\alpha,k}=\sum_{i=1}^{N_k}\hat{\sigma}^{(i)}_{\alpha,k}/2$ is a collective spin operator on mode $k$ with Pauli matrices $\hat{\sigma}^{(i)}_{\alpha,k}$ for $\alpha=x,y,z$
and $k=1,..., M$. Without loss of generality, we label the axes such that the mean spin direction $\mathbf{n}_{0,k}=\langle\hat{\mathbf{J}}_k\rangle_{\hat{\rho}}/|\langle\hat{\mathbf{J}}_k\rangle_{\hat{\rho}}|$ defines the $z$ axis. By considering a family of local measurement observables $\hat{\mathbf{X}}=\hat{\mathbf{J}}_{\mathbf{s}}$, we obtain the spin-squeezing matrix with elements
\begin{align}\label{eq:winelandmatrix}
(\matxi^2[\hat{\rho},\hat{\mathbf{J}}_{\mathbf{r}},\hat{\mathbf{J}}_{\mathbf{s}}])_{kl}=\frac{\sqrt{N_kN_l}\mathrm{Cov}(\hat{J}_{\mathbf{s}_k,k},\hat{J}_{\mathbf{s}_l,l})_{\hat{\rho}}}{\langle\hat{J}_{z,k}\rangle_{\hat{\rho}}\langle\hat{J}_{z,l}\rangle_{\hat{\rho}}},
\end{align}
where we used Eq.~(\ref{eq:squeezingmatrix}) with $\matF_{\mathrm{SN}}[\hat{\mathbf{J}}_{\mathbf{r}}]^{\frac{1}{2}}\matC[\hat{\rho},\hat{\mathbf{J}}_{\mathbf{r}},\hat{\mathbf{J}}_{\mathbf{s}}]^{-1}=\mathrm{diag}(\sqrt{N_1}/\langle\hat{J}_{z,1}\rangle_{\hat{\rho}},\dots,\sqrt{N_M}/\langle\hat{J}_{z,M}\rangle_{\hat{\rho}})$ and we assumed that the $\mathbf{r}_k$ and $\mathbf{s}_k$ are orthonormal vectors in the $xy$-plane, such that $\langle\hat{J}_{z,k}\rangle_{\hat{\rho}}=-i\langle[\hat{J}_{\mathbf{s}_k,k},\hat{J}_{\mathbf{r}_k,k}]\rangle_{\hat{\rho}}$ is the length of spin $k$ with mean-spin direction along the $z$ axis. 
On its diagonal, this matrix contains the local spin-squeezing coefficients \cite{Wineland}, for each of the modes $k=1,\dots,M$. It is well known that these coefficients reveal the number of entangled spins within the local modes~\cite{RMP,SMPRL01,Sorensen}. In addition to these single-parameter contributions, the multiparameter spin-squeezing matrix~(\ref{eq:winelandmatrix}) includes off-diagonal terms that are due to mode correlations, i.e., entanglement between the individual interferometers.

{\bf Atomic multiparameter spin squeezing.}
A locally squeezed state can be created by subjecting spatially separated ensembles of atoms to local, nonlinear evolutions, e.g., by means of the one-axis twisting Hamiltonian \cite{KitagawaUeda}. It is easy to see from the squeezing matrix that local squeezing is sufficient to attain full multiparameter sub-shot-noise; see \supp{3} for details. However, atomic experiments are not limited to the generation of local squeezing: Recently, spatially distributed entanglement was observed by splitting squeezed atomic spin ensembles into two or more external modes \cite{FadelSCIENCE2018,KunkelSCIENCE2018,LangeSCIENCE2018}.

In order to identify the metrological potential of nonlocal squeezing, we compare two different spin squeezing strategies.
We consider an even number of $N$ spin-1/2 particles initialized in the polarized state $|\Psi_0\rangle = \vert \uparrow \rangle^{\otimes N}$, where $|\uparrow\rangle$ is an eigenstate of the Pauli $z$ matrix. Local squeezing 
(namely, local in each atomic ensemble) corresponds to 
\begin{align} \label{Psiloc}
|\Psi_{\rm loc}(t)\rangle=e^{-i(\hat{J}_{y,1}^2+\hat{J}_{y,2}^2)\chi t}|\Psi_0\rangle, 
\end{align}
where $\hat{J}_{y,1}$ and $\hat{J}_{y,2}$ are collective spin operators for particles $1,2,\dots, N/2$ and $N/2+1, \dots, N$, respectively, i.e., we have separated the particles into two ensembles of equal size.
The nonlinear evolution generates entanglement between the $N/2$ particles in each ensemble, e.g., by describing interactions among the particles in the same ensemble for the dimensionless time $\chi t$, but does not entangle the two ensembles.
Nonlocal squeezing is instead described by the collective one-axis-twisting evolution 
\begin{align} \label{Psinonloc}
|\Psi_{\rm nl}(t)\rangle=e^{-i(\hat{J}_{y,1}+\hat{J}_{y,2})^2\chi t}|\Psi_0\rangle
\end{align}
which creates particle entanglement between the $N$ spins and mode entanglement between the two ensembles.

Our goal is to estimate linear combinations $\mathbf{n}^T\vectheta=n_1\theta_1+n_2\theta_2$ of locally encoded parameters, generated by the rotations $\hat{J}_{\mathbf{r}_1,1}$ and $\hat{J}_{\mathbf{r}_2,2}$ via the transformation $\hat{U}(\vectheta)=\exp(-i\hat{J}_{\mathbf{r}_1,1}\theta_1-i\hat{J}_{\mathbf{r}_2,2}\theta_2)$. A particular case of interest is the estimation of a magnetic field gradient \cite{Koschorreck,UrizarPRA2013,AltenburgPRA2017,ApellanizPRA2018} based on the differential measurement of the field at two spatially separated locations, which corresponds to the difference $\mathbf{n}_{-}=(1,-1)^T/\sqrt{2}$. A related task is the estimation of the average field, i.e., the sum of parameters $\mathbf{n}_{+}=(1,1)^T/\sqrt{2}$. We assume that the local rotation axes $\mathbf{r}_1$ and $\mathbf{r}_2$ (and their corresponding optimal measurement directions) can be adjusted to optimize the local squeezing parameters, i.e., to minimize the diagonal entries of the squeezing matrix~(\ref{eq:winelandmatrix}). Such a change of the rotation axis can effectively be realized through local rotations of the respective spin states before the interferometric measurement \cite{RMP}.

\begin{figure}[tb]
\centering
\includegraphics[width=.49\textwidth]{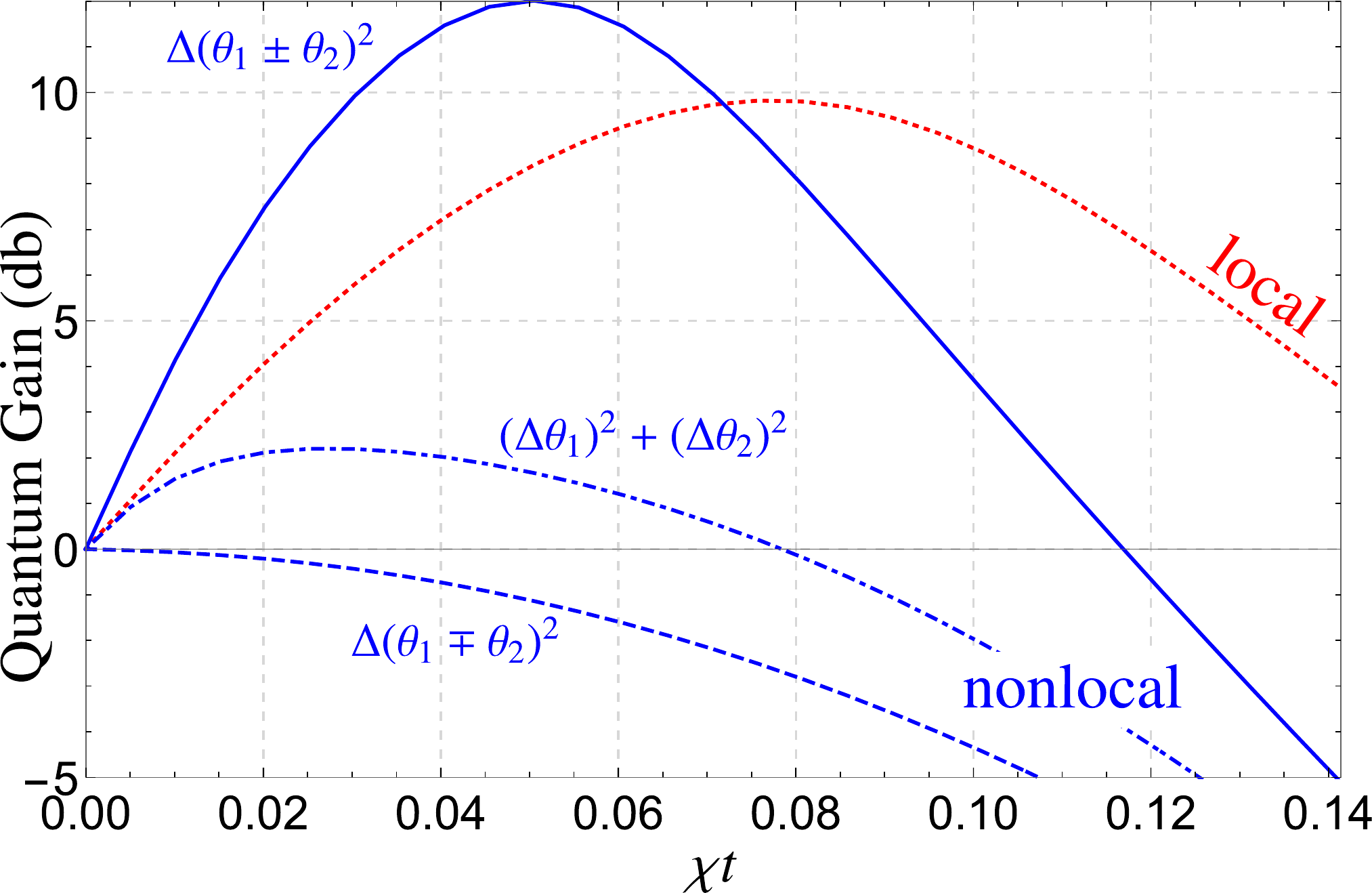}
\caption{Local vs nonlocal atomic spin squeezing. For a local parameter encoding with $N=100$ particles, nonlocal squeezing, described by Eq.~(\ref{Psiloc}), leads to a larger quantum sensitivity gain for either the sum $10\log_{10}(\mathbf{n}_+^T\matSigma_{\rm SN}\mathbf{n}_+/\mathbf{n}_+^T\matSigma\mathbf{n}_+)$ (continuous blue line) or the difference of two spatially distributed parameters $10\log_{10}(\mathbf{n}_-^T\matSigma_{\rm SN}\mathbf{n}_-/\mathbf{n}_-^T\matSigma\mathbf{n}_-)$ (dashed blue line) than local squeezing, Eq.~(\ref{Psinonloc}). Since the spin-squeezing matrix is diagonal when squeezing is local, both combinations of parameters, as well as their uncorrelated average yield the same sensitivity (red dashed line). Nonlocal squeezing yields a lower quantum gain for the uncorrelated average $10\log_{10}(\mathrm{Tr}\matSigma_{\rm SN}/\mathrm{Tr}\matSigma)$ (dash-dotted line). The plot shows data for local directions $\mathbf{r}_1$ and $\mathbf{r}_2$ chosen to maximize the gain for the sum. A local rotation transforms the sum of parameters into the difference and vice-versa.}
\label{fig:nlsqz}
\end{figure}

The resulting sensitivities for $\mathbf{n}_{\pm}^T\vectheta$ are compared in Fig.~\ref{fig:nlsqz} for an ensemble of $N=100$ atoms as a function of the nonlinear evolution time $t$. We observe that an estimation of, e.g., $\mathbf{n}_+^T\vectheta$ can be enhanced by nonlocal squeezing (blue continuous line). As a consequence, the sensitivity for $\mathbf{n}_-^T\vectheta$ is reduced below the classical limit (blue dashed line). However, a local $\pi$-rotation of the state can effectively change the sign of $\mathbf{r}_2$ and transform the sum into the difference and vice-versa. Hence, nonlocal squeezing can be used to reduce the uncertainty of a specific linear combination of parameters. The state cannot be optimal for arbitrary linear combinations at the same time, but local operations can be used to adjust the state prior to the measurement in order to optimally harness the nonlocal squeezing and beat the sensitivity of local squeezing. Nonlocal squeezing further improves the estimation of nonlocally encoded parameters, as we discuss in \supp{3}. 

{\bf Multiparameter continuous-variable squeezing.} 
Continuous-variable multiparameter estimation studies the sensitivity to a multimode displacement described by phase space operators $\hat{\mathbf{q}}=(\hat{x}_1,\hat{p}_1,\dots,\hat{x}_M,\hat{p}_M)^T$, where $\hat{x_k}=\frac{1}{2}(\hat{a}_k+\hat{a}_k^{\dagger})$ and $\hat{p}_k=\frac{1}{2i}(\hat{a}_k-\hat{a}_k^{\dagger})$ and $[\hat{a}_k,\hat{a}_{k'}^{\dagger}]=\delta_{kk'}$. 
These observables are accessible by homodyne measurement techniques, i.e., by mixing the signal with a strongly populated local oscillator---a well established technique in optical  \cite{BraunsteinVanLoock,Ferraro,Wang,Weedbrook,YokoyamaNATPHOT2013,CaiNATCOMM2017} and atomic systems~\cite{GrossNATURE,EPRatomic}.

The $2M\times 2M$ moment matrix, Eq.~(\ref{eq:momentmatrixA}), for $\hat{\mathbf{A}} = \hat{\mathbf{q}}$ reads
\begin{align}\label{eq:CVM}
\tilde{{\matM}}[\hat{\rho},\hat{\mathbf{q}}]=\frac{1}{4}\matOmega^T\matGamma[\hat{\rho},\hat{\mathbf{q}}]^{-1}\matOmega,
\end{align}
and provides the maximally achievable sensitivity for multimode displacements via Eq.~(\ref{eq:maxM}). The $2M\times 2M$ covariance matrix $\matGamma[\hat{\rho},\hat{\mathbf{q}}]$ contains complete information on non-displaced Gaussian states. The commutator matrix $\tilde{\matC}[\hat{\rho},\hat{\mathbf{q}}]=\frac{1}{2}\matOmega$ is independent of the quantum state, where $\matOmega=\bigoplus_{k=1}^M\matomega$ is the symplectic form with $\matomega=\left(\begin{smallmatrix} 0 & 1\\-1 & 0 \end{smallmatrix}\right)$ \cite{Ferraro,Wang,Weedbrook}. Furthermore, the explicit evaluation of the quantum Fisher matrix of Gaussian states  $\hat{\rho}_{\rm G}$ \cite{Monras,Pinel13,Zhang2014} (i.e. states whose Wigner function is Gaussian \cite{Ferraro,Wang,Weedbrook}) reveals that it coincides with Eq.~(\ref{eq:CVM}). We thus obtain the exact equality $\tilde{{\matM}}[\hat{\rho}_{\rm G},\hat{\mathbf{q}}]=\matF_Q[\hat{\rho}_{\rm G},\hat{\mathbf{q}}]$ for arbitrary Gaussian states $\hat{\rho}_{\rm G}$, whereas for arbitrary quantum states $\hat{\rho}$, Eq.~(\ref{eq:CVM}) represents a Gaussian lower bound to the quantum Fisher matrix, see Eq.~(\ref{eq:MF}). Making use of upper bounds on the quantum Fisher matrix for specific classes of separable states~\cite{GessnerPRA2016,Quantum2017,GessnerPRL2018}, the moment matrix can reveal detailed information about the multimode entanglement structure~\cite{QinNPJQI2019}.

The continuous-variable squeezing matrix, optimized over the measurement observables $\hat{\mathbf{X}}$, is given by Eq.~(\ref{eq:xiopt}) and reads:
\begin{align}\label{eq:CVxiopt}
\matxi_{\mathrm{opt}}^{2}[\hat{\rho},\hat{\mathbf{H}},\hat{\mathbf{q}}]=4\matR\matOmega^T\matGamma[\hat{\rho},\hat{\mathbf{q}}]\matOmega\matR^T.
\end{align}
Let us first revisit the general squeezing condition for the particular case of the multimode continuous-variable system at hand. A violation of~(\ref{eq:SQZcond}) implies that (see \supp{4})
\begin{align}\label{eq:sqzSimon}
\lambda_{\min}(\matGamma[\hat{\rho},\hat{\mathbf{q}}])<\frac{1}{4},
\end{align}
where $\lambda_{\min}$ denotes the smallest eigenvalue. 
The condition~(\ref{eq:sqzSimon}) was originally proposed in Ref.~\cite{SimonPRA1994} as a definition of squeezing in multimode continuous-variable systems that is invariant under passive transformations, i.e., beam splitter operations and phase shifters that leave the number of photons constant. Conversely, if~(\ref{eq:sqzSimon}) holds, one can find $\hat{\mathbf{H}}$ and $\hat{\mathbf{X}}$ such that the condition~(\ref{eq:SQZcond}) is violated. 

Hence, our general metrological definition of squeezing in multimode systems is equivalent to a well-established definition~\cite{SimonPRA1994} in the continuous-variable case when considering quadrature operators. The shot-noise rank $r_{\rm SN}$, i.e., the number of eigenvalues of $\matxi_{\mathrm{opt}}^{2}[\hat{\rho},\hat{\mathbf{H}},\hat{\mathbf{q}}]$ that are smaller than one, provides a step-wise characterization of the multiparameter quantum gain up to full multiparameter squeezing (namely $r_{\rm SN}=M$). This establishes a natural multiparameter extension of the single-parameter condition~(\ref{eq:sqzSimon}), which merely implies that $r_{\rm SN}>0$. 

\begin{figure}[tb]
\centering
\includegraphics[width=.4\textwidth]{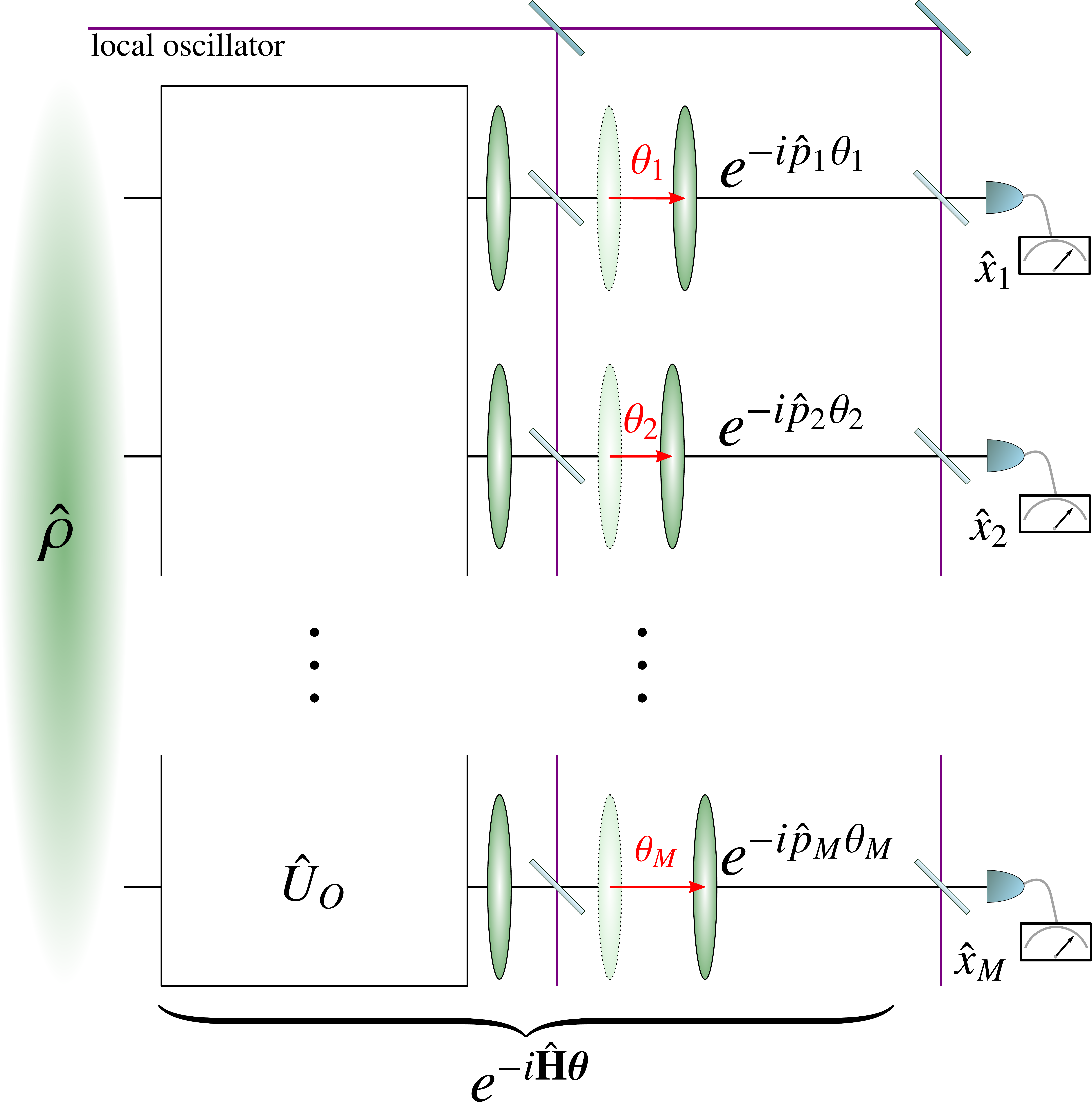}
\caption{Optimal multimode displacement sensing with squeezed vacuum states. The passive transformation $\hat{U}_{\matO}$ decouples the initial multimode squeezed vacuum state into local squeezed states. A displacement generated by the anti-squeezed variance (here depicted as $\hat{p}$) and a measurement of the squeezed variance ($\hat{x}$) is implemented in each mode with the aid of a local oscillator.} \label{CVMMZ}
\end{figure}

{\bf Multimode squeezed vacuum states.}
The class of pure Gaussian continuous-variable states is given by multimode squeezed vacuum states $|\Psi_0\rangle$ \cite{BraunsteinVanLoock,Ferraro,Wang,Weedbrook}. 
As a consequence of the Williamson theorem and the Bloch-Messiah decomposition \cite{BraunsteinPRA2005}, any such state can be generated by a combination of local squeezing and a series of passive operations \cite{Ferraro,Weedbrook}. Consequently, there always exists a $2M\times 2M$ orthogonal symplectic matrix $\matO$, and a corresponding passive operation described by $\hat{U}_{\matO}$, that yields $\matGamma[\hat{U}_{\matO}|\Psi_0\rangle,\hat{\mathbf{q}}]=\matO\matGamma[|\Psi_0\rangle,\hat{\mathbf{q}}]\matO^T=\frac{1}{4}\bigoplus_{k=1}^M\mathrm{diag}(e^{2r_k},e^{-2r_k})$, where $r_1,\dots,r_M$ quantify the squeezing in each of the modes. 

The choice of phase-encoding Hamiltonians and measurement observables 
$\hat{\mathbf{H}}=\matU\matP_M\matO\matOmega\hat{\mathbf{q}}$ and $\hat{\mathbf{X}}=\matP_M\matO\hat{\mathbf{q}}$, where $\matU$ is an arbitrary $M\times M$ orthogonal matrix and $\matP_M$ is a $M\times 2M$ projector that picks one quadrature per mode, is optimal (see Methods and \supp{4} for details) and leads to
\begin{align}\label{eq:mmSQZ}
\matxi_{\mathrm{opt}}^{2}[|\Psi_0\rangle,\hat{\mathbf{H}},\hat{\mathbf{q}}]=\matU\begin{pmatrix} e^{-2r_1}  & \dots & 0\\
\vdots & \ddots & \vdots\\
0 & \cdots  & e^{-2r_M}
\end{pmatrix}\matU^T.
\end{align}
These operators can be interpreted (see Fig.~\ref{CVMMZ}) as a phase-imprinting evolution that first disentangles the state and then implements local phase shifts along the respective squeezed quadrature in each mode. The measurement is realized in the corresponding conjugate quadrature. If all $r_k>0$, we have full multiparameter squeezing, enabling sub-shot-noise estimation of arbitrary linear combinations of the parameters $\vectheta$ encoded via the evolution $U(\vectheta)=e^{-i(\hat{H}_{1}\theta_1+\cdots+\hat{H}_{M}\theta_M)}$.

For $\matO=\matID_{2M}$ and $\matU=\matID_M$, the parameter encoding realized by the Hamiltonians $\hat{\mathbf{H}}$ is local in the modes $\hat{\mathbf{q}}$. The result~(\ref{eq:mmSQZ}) shows that the multiparameter sensitivity of local transformations is maximized by a mode-local product state. Similarly, for any other choice of $\matO$, we can define new modes $\matO\hat{\mathbf{q}}$ as nonlocal linear combinations of the original $\hat{\mathbf{q}}$, and for transformations that are local in $\matO\hat{\mathbf{q}}$, the sensitivity is maximized by states that are uncorrelated in the modes $\matO\hat{\mathbf{q}}$. These states will generally be mode entangled in the original set of modes $\hat{\mathbf{q}}$. We conclude that mode entanglement with respect to the modes $\hat{\mathbf{q}}$ is not necessary to optimize the overall multiparameter sensitivity if the parameter encoding is done locally in $\hat{\mathbf{q}}$. Conversely, given a transformation that is nonlocal in $\hat{\mathbf{q}}$, the optimal sensitivity is achieved by a mode entangled state.

{\bf Maximum enhancement due to mode entanglement.}
Recall that the multiparameter covariance matrix contains information equivalent to the sensitivity of arbitrary linear combinations of parameters. For any specific linear combination, local squeezing is still suboptimal (an analogue observation was discussed above for the case of spins). In this case, we are interested in minimizing a single matrix element rather than all eigenvalues of the squeezing matrix. Let us now identify the maximum gain that can be achieved by making use of mode entanglement.

\begin{figure}[tb]
\centering
\includegraphics[width=.45\textwidth]{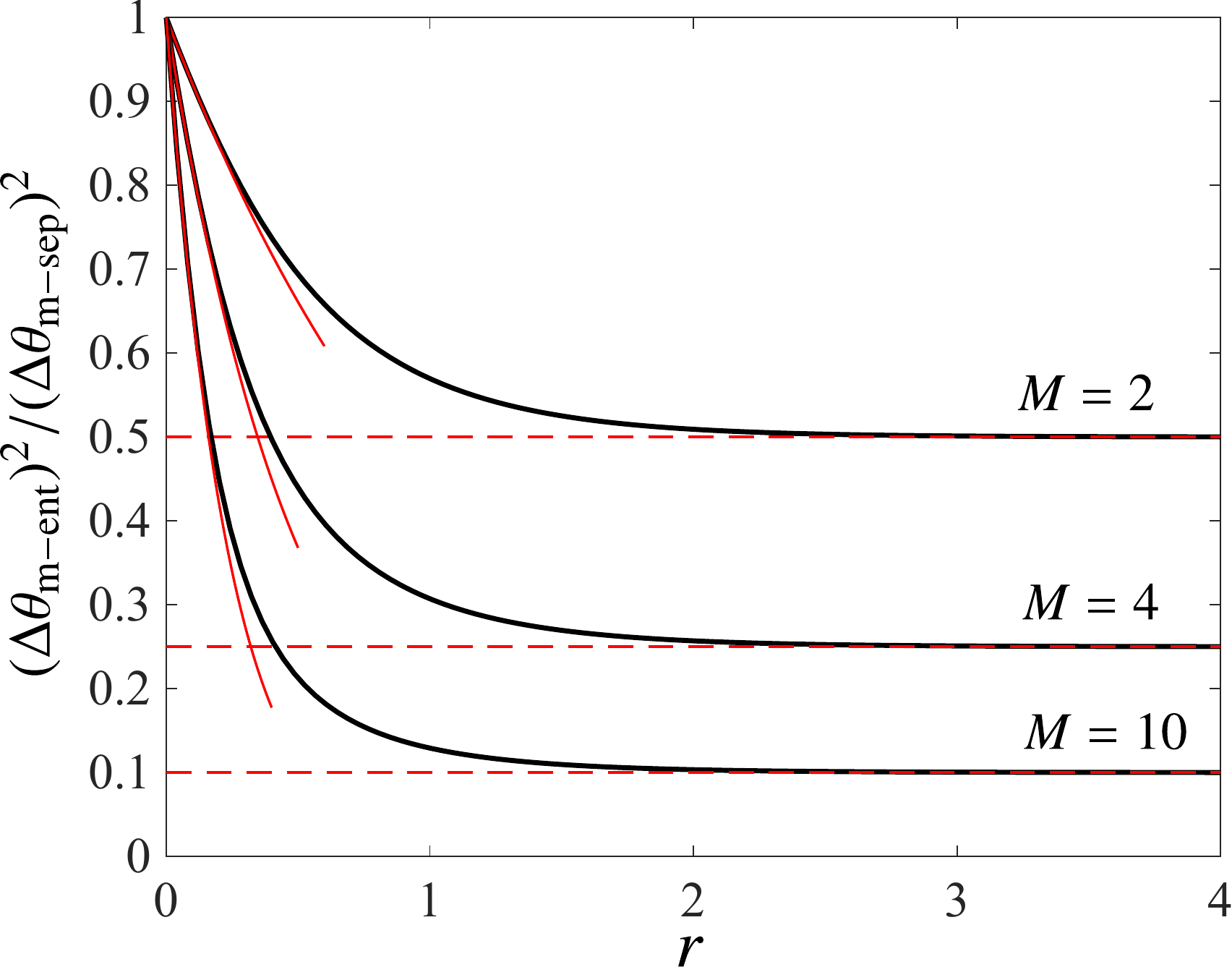}
\caption{Quantum gain from nonlocal mode entanglement. We plot the ratio between the sensitivity to an uniform average of parameters $(\Delta \theta)^2 = \mathbf{n}^T\mathbf{\Sigma} \mathbf{n}$
for optimal mode-entangled and mode-separable states (thick black line), as a function of the squeezing parameter $r$.
The solid red lines are the small-$r$ approximation $e^{-2 (\sqrt{M}-1) r}$ and the dashed red lines are the large-$r$ approximation $1/M$. Different sets of lines refer to different values of $M$.}
\label{fig:sqCV}
\end{figure}

We consider a fixed family of phase-imprinting Hamiltonians (hence $\matU=\matID_{M}$) and an estimation of $\mathbf{n}^T\vectheta$ with an arbitrary, fixed unit vector $\mathbf{n}$ that has non-zero overlap with all the participating modes $k=1,\dots,M$. Our goal is to distribute a finite total amount of squeezing (determined by the total average particle number) over all modes in order to minimize $\mu\mathbf{n}^T\matSigma\mathbf{n}=\mathbf{n}^T\matxi^2_{\mathrm{opt}}[|\Psi_0\rangle,\hat{\mathbf{H}}, \mathbf{q}]\mathbf{n}$. We compare the optimized mode-separable and mode-entangled strategy (see Methods for details), giving rise to the respective sensitivities $(\Delta \theta_{\rm m-sep})^2$ and $(\Delta \theta_{\rm m-ent})^2$: For a uniform average over all parameters, $n_k=1/\sqrt{M}$, the optimal mode-separable strategy consists in equal squeezing in all modes, $r_k=r$, for $k=1,\dots,M$, while the optimal mode-entangled strategy concentrates all squeezing into a single mode. As soon as $r>0$, we have $(\Delta \theta_{\rm m-ent})^2 / (\Delta \theta_{\rm m-sep})^2 < 1$, see Fig.~\ref{fig:sqCV}: the mode-entangled strategy outperforms the mode-separable one. In the limit $r \ll 1/\sqrt{M}$ we  obtain $(\Delta \theta_{\rm m-ent})^2 / (\Delta \theta_{\rm m-sep})^2 \approx e^{-2 (\sqrt{M}-1) r}$.
In the opposite limit, $r \gg1$, we have $e^{-2 r'} \approx M e^{-2r}$ and we obtain
\begin{align}
\frac{(\Delta \theta_{\rm m-ent})^2 }{ (\Delta \theta_{\rm m-sep})^2} = \frac{1}{M}\qquad (r\gg 1).
\end{align}
We thus recover the gain factor $1/M$ that has been identified as the maximal gain due to mode entanglement~\cite{HumphreysPRL2013,ProctorPRL2018,GePRL2018,GessnerPRL2018,Guo2019}. Here, the factor $1/M$ is obtained by comparing optimal Gaussian states based on the analysis of the multimode squeezing matrix. We further show in \supp{4} that among all possible states with fixed average particle number, squeezed vacuum states optimize the sensitivity of multiparameter displacement sensing, generalizing the single-parameter results of Refs.~\cite{Lang,MatsubaraNJP2019}.

{\bf Non-commuting generators and non-Gaussian states.}
To illustrate how our methods can lead to efficient and saturable strategies in more general scenarios, we now discuss an example dedicated to the estimation of parameters that are generated by non-commuting operators using a non-Gaussian state.

We consider the estimation of the two angles $\theta_{1,2}$ of a $\mathrm{SU}(2)$ rotation
$\hat{U}(\vectheta) = e^{-i (\theta_1 \hat{J}_x + \theta_2 \hat{J}_y)}$ with non-commuting generators $\hat{\mathbf{H}}=(\hat{J}_x,\hat{J}_y)^T$ in a single mode. As probe, we use the twin-Fock state $|\mathrm{TF}\rangle$, i.e., eigenstates of $\hat{J}_z$ with eigenvalue zero and a total spin length of $N/2$, and we denote $\vert {\rm TF}_{\vectheta} \rangle = \hat{U}(\vectheta) \vert {\rm TF} \rangle$. Having zero mean spin length, 
$|\mathrm{TF}\rangle$ cannot be characterized by spin squeezing~\cite{TF} and Gaussian measurements are unable to fully harness its metrological potential. We consider the two commuting nonlinear observables $\hat{X}_1 = \hat{J}_x |\mathrm{TF}\rangle \langle \mathrm{TF} \vert \hat{J}_x$
and $\hat{X}_2 = \hat{J}_y |\mathrm{TF}\rangle \langle \mathrm{TF} \vert \hat{J}_y$:
as a consequence of $\langle \mathrm{TF} \vert \hat{J}_x \hat{J}_y \vert \mathrm{TF} \rangle = 0$,
we have $[\hat{X}_1, \hat{X}_2] = 0$.
Let us indicate with $Q = \langle \mathrm{TF} \vert \hat{J}_x^2 \vert \mathrm{TF} \rangle = \langle \mathrm{TF} \vert \hat{J}_y^2 \vert \mathrm{TF} \rangle = \tfrac{N(N+2)}{8}$.
To the leading order in $\theta_{1,2}$, we obtain the inverse covariance matrix $\matGamma[\vert {\rm TF}_{\vectheta} \rangle,\hat{\mathbf{X}}]^{-1} =Q^{-3}\mathrm{diag}(\theta_1^{-2},\theta_2^{-2})$ and the commutator matrix $\matC[\vert {\rm TF}_{\vectheta} \rangle,\hat{\mathbf{H}},\hat{\mathbf{X}}] =2Q^2 \mathrm{diag}(\theta_1,\theta_2)$. In the limit $\vectheta\to \bs{0}$, this leads to the moment matrix~(\ref{eq:momentmatrix}) 
\begin{align}\label{eq:TFQFI}
\matM[|\mathrm{TF}\rangle,\hat{\mathbf{H}},\hat{\mathbf{X}}]=\frac{N(N+2)}{2}\matID_2,
\end{align}
which coincides with the quantum Fisher matrix $\matF_Q[|\mathrm{TF}\rangle,\hat{\mathbf{H}}]$. This shows that through the measurement of nonlinear observables, our method can extract the full sensitivity of non-Gaussian states, and that it can achieve the ultimate multiparameter sensitivity limit even when the generators do not commute.

\section*{Discussion} 
We introduced metrological multiparameter squeezing as a practical framework to characterize the sensitivity and quantum gain of multiparameter estimation. Our optimization technique can be adapted to any set of accessible observables and thereby allows to adjust the level of complexity to the problem at hand. 
For example, the multiparameter sensitivity of Gaussian states can be fully captured by a squeezing matrix 
only containing first and second moments of linear observables. The analysis of the squeezing matrix reveals optimal strategies for the design and analysis of atomic and photonic experiments where Gaussian states still represent the best-controlled and most efficiently generated class of states for metrology. Metrological multiparameter squeezing thus lays the foundation for the development of atomic clocks and electromagnetic field sensors, enhanced by non-local quantum correlations in atomic ensembles with spatially distributed and accessible entanglement~\cite{GreinerNATURE2009,LabuhnNATURE2016,LanyonPRX2018,LukinSCIENCE2019,FadelSCIENCE2018,KunkelSCIENCE2018,LangeSCIENCE2018,Jing2019}. Furthermore, optical systems provide an established platform with access to entangled multimode photonic quantum states \cite{YokoyamaNATPHOT2013,CaiNATCOMM2017,PolinoOPTICA2019} that can be combined with squeezing \cite{Steinlechner,TrepsSCIENCE2003}. Our theory of multiparameter squeezing provides a common framework to characterize these experiments and to interpret and optimize them for multiparameter quantum sensing applications. 

By extending the set of accessible observables, the squeezing matrix can be generalized to yield more powerful quantifiers of multiparameter sensitivity that are able to cope with highly sensitive features of non-Gaussian multimode states. This method can also be applied in non-commuting scenarios, where, however, further studies are needed to explore the full potential of our approach. Such developments are important, e.g., in optical systems where one aims to estimate the coordinates of an ensemble of emitters to reconstruct an image~\cite{KolobovBOOK,TrepsSCIENCE2003,TsangPRX2016,TsangPRL2016,LupoPRL2016,RehacekPRA2017}. The identification of fundamental resolution limits for quantum imaging requires experimentally and theoretically accessible measures of multiparameter sensitivity for arbitrary emitters.

\section*{Methods}
\textbf{Multiparameter method of moments.} 
We base our multiparameter method of moments on the knowledge of the mean
values of a family of commuting observables, $\langle \hat{\mathbf{X}}\rangle_{\hat{\rho}(\vectheta)}=(\langle\hat{X}_1\rangle_{\hat{\rho}(\vectheta)},\dots,\langle\hat{X}_K\rangle_{\hat{\rho}(\vectheta)})^T$, 
obtained from the calibration of the experimental apparatus
as a function of the $M$ parameters $\vectheta=(\theta_1,\dots,\theta_M)^T$.
If $\hat{\mathbf{X}}$ is measured $\mu\gg 1$ times, each of its components  $\hat{X}_k$ yields a sequence of results $x^{(1)}_k, \dots, x^{(\mu)}_k$ where the $x^{(i)}_k$ are picked from the eigenvalues of $\hat{X}_k$. Each measurement of $\hat{\mathbf{X}}$ thus yields a vector of results $\mathbf{x}^{(i)}=(x_1^{(i)}, \dots, x_K^{(i)})^T$ that is randomly distributed with mean value $\langle \hat{\mathbf{X}}\rangle_{\hat{\rho}(\vectheta)}$ and covariance matrix 
$(\matGamma[\hat{\rho}(\vectheta),\hat{\mathbf{X}}])_{kl}=\langle\hat{X}_k\hat{X}_l\rangle_{\hat{\rho}(\vectheta)}-\langle\hat{X}_k\rangle_{\hat{\rho}(\vectheta)}\langle\hat{X}_l\rangle_{\hat{\rho}(\vectheta)}$. From these measurements we obtain the sample average $\bar{\mathbf{X}}^{(\mu)}=(\bar{X}^{(\mu)}_1,\dots,\bar{X}^{(\mu)}_K)^T$ with $\bar{X}^{(\mu)}_k=\frac{1}{\mu}\sum_{i=1}^\mu x^{(i)}_k$ for $k=1,\dots,K$. We estimate the parameters $\vectheta$ as the values for which $\langle\hat{X}_k\rangle_{\hat{\rho}(\vectheta)}=\bar{X}_k^{(\mu)}$ holds for all $k=1,\dots,K$. As a consequence of the multivariate central limit theorem (see \supp{1} for details), for $\mu\gg 1$, this strategy yields $\matSigma=(\mu \matM[\hat{\rho}(\vectheta),\hat{\mathbf{X}}])^{-1}$, where
\begin{align}\label{eq:MmatrixD}
\matM[\hat{\rho}(\vectheta),\hat{\mathbf{X}}]=\matD[\hat{\rho}(\vectheta),\hat{\mathbf{X}}]^T\matGamma[\hat{\rho}(\vectheta),\hat{\mathbf{X}}]^{-1}\matD[\hat{\rho}(\vectheta),\hat{\mathbf{X}}],
\end{align}
and
\begin{align}\label{eq:dmatrix}
(\matD[\hat{\rho}(\vectheta),\hat{\mathbf{X}}])_{kl}=\frac{\partial\langle\hat{X}_k\rangle_{\hat{\rho}(\vectheta)}}{\partial \theta_l}.
\end{align}
In the case of a single parameter estimated by a single observable ($M=K=1$) we obtain a sensitivity described by the familiar error propagation formula 
$(\Delta \theta_{\mathrm{est}})^2=\frac{1}{\mu}(\Delta \hat{H})_{\hat{\rho}(\theta)}^2
\left|\frac{\partial\langle \hat{X}\rangle_{\hat{\rho}(\theta)}}{\partial \theta}\right|^{-2}$, where $\frac{\partial\langle \hat{X}\rangle_{\hat{\rho}(\theta)}}{\partial \theta}=-i\langle [\hat{X},\hat{H}]\rangle_{\hat{\rho}(\theta)}$ \cite{Wineland}. The result~(\ref{eq:MmatrixD}) provides a direct generalization to the multiparameter case.

For a unitary phase imprinting processes $U(\vectheta)=\exp(-i\hat{\mathbf{H}}\vectheta)$, generated by the vector of Hamiltonians $\hat{\mathbf{H}}=(\hat{H}_1,\dots,\hat{H}_M)^T$, as considered in the main text, we obtain
\begin{align}
(\matD[\hat{\rho}(\vectheta),\hat{\mathbf{X}}])_{kl}
=-i\langle [\hat{X}_k,\hat{H}_l]\rangle_{\hat{\rho}(\vectheta)}=(\matC[\hat{\rho}(\vectheta),\hat{\mathbf{H}},\hat{\mathbf{X}}])_{kl},
\end{align}
and we recover the moment matrix given in Eq.~(\ref{eq:gaussianfishermatrix}). We have assumed that $\matGamma[\hat{\rho}(\vectheta),\hat{\mathbf{X}}]$ is invertible and that $\matD[\hat{\rho}(\vectheta),\hat{\mathbf{X}}]$ has rank $M$. This is usually the case for a suitable choice of the operators $\hat{\mathbf{X}}$ as is illustrated by our application to relevant examples of spin and continuous-variable systems. Rank-deficiency of these matrices may indicate a redundancy in the information provided by the vector of measurement results that can be remedied by reducing the number of observables.

\textbf{Multiparameter shot-noise limit.} The classical precision limit of multiparameter distributed sensor networks, i.e., the multiparameter shot-noise limit, is defined as the maximal sensitivity that can be achieved by some optimally chosen classical probe state \cite{GessnerPRL2018},
\begin{align}\label{eq:MSN}
\matF_{\mathrm{SN}}[\hat{\mathbf{H}}]:=\max_{\hat{\rho}_{\rm cl}}\matF_Q[\hat{\rho}_{\rm cl},\hat{\mathbf{H}}].
\end{align}
The family of classical probe states $\hat{\rho}_{\rm cl}$ depends on the system at hand. For a fixed number of particles, the system can effectively be described by discrete variables and a natural definition of classical states is given by particle-separable states \cite{PS09,RMP}. Similarly, for continuous-variable systems we consider mixtures of coherent states as classical \cite{WallsBOOK,RivasPRL2010}. In the single-parameter theory, these families of classical states yield familiar expressions for the shot-noise limit, i.e., the $1/N$-scaling of the variance when $N$ is the number of particles, or the uncertainty of the vacuum state for homodyne measurements. These limits can be generalized to the multiparameter case, where the shot-noise matrix~(\ref{eq:MSN}) is diagonal for locally encoded parameters~\cite{GessnerPRL2018}. The shot-noise limit for evolutions generated by $\hat{\mathbf{H}}$ is obtained from the quantum Cram\'er-Rao bound $\matSigma\geq(\mu \matF_Q[\hat{\rho},\hat{\mathbf{H}}])^{-1}$ by considering the sensitivity of the optimal classical state:
\begin{align}\label{eq:SNL}
\matSigma_{\mathrm{SN}}=(\mu\matF_{\mathrm{SN}}[\hat{\mathbf{H}}])^{-1}.
\end{align}
As a consequence of Eq.~(\ref{eq:MF}), we obtain that Eq.~(\ref{eq:SQZcond}) holds for all classical states $\rho_{\mathrm{cl}}$.

The shot-noise limit in discrete-variable multimode interferometers is attained by the most sensitive particle-separable state 
$\hat{\rho}_{\mathrm{p-sep}}=\sum_{\gamma}p_{\gamma}\hat{\rho}^{(\gamma)}_1\otimes\cdots\otimes\hat{\rho}^{(\gamma)}_N$, where $p_{\gamma}$ is a probability distribution and the $\hat{\rho}^{(\gamma)}_k$ are quantum states of particle $k$.
Optimization over separable states leads to the shot-noise limit~(\ref{eq:SNL}) defined in terms of a diagonal quantum Fisher matrix with diagonal elements given by the respective numbers of particles in each mode \cite{GessnerPRL2018}.
Specifically, the classical sensitivity limit as a function of the accessible operators $\hat{\mathbf{J}}_{\perp}$ reads $\matF_{\mathrm{SN}}[\hat{\mathbf{J}}_{\perp}]:=\max_{\hat{\rho}_{\mathrm{p-sep}}}\matF_Q[\hat{\rho}_{\mathrm{p-sep}},\hat{\mathbf{J}}_{\perp}]=\mathrm{diag}(N_1,N_1,\dots,N_M,N_M)$. For Hamiltonians $\hat{\mathbf{H}}=\matR\hat{\mathbf{J}}_{\perp}$ [recall Eq.~(\ref{eq:HXRS})] that consist of linear combinations of the elements of $\hat{\mathbf{J}}_{\perp}$ this implies $\matF_{\mathrm{SN}}[\hat{\mathbf{H}}]=\matR\matF_{\mathrm{SN}}[\hat{\mathbf{J}}_{\perp}]\matR^T$. Further details are provided in \supp{3}. Sensitivities beyond this limit can be achieved only by employing particle entanglement.

\textbf{Optimization of the phase-imprinting Hamiltonians.} To optimize the choice of $\matR$, i.e., the phase-imprinting Hamiltonians $\hat{\mathbf{H}}$, recall that the optimal moment matrix $\matM_{\rm opt}[\hat{\rho},\hat{\mathbf{H}},\hat{\mathbf{A}}]$ describes an $M\times M$ orthogonal projection of the larger $L\times L$ matrix $\tilde{\matM}[\hat{\rho},\hat{\mathbf{A}}]$. The eigenvectors and eigenvalues of $\matM_{\rm opt}[\hat{\rho},\hat{\mathbf{H}},\hat{\mathbf{A}}]$ both depend on the $M\times L$ matrix $\matR$. First, we notice that the basis of $\matM_{\rm opt}[\hat{\rho},\hat{\mathbf{H}},\hat{\mathbf{A}}]$ can be chosen at will by orthogonal transformations of the generating Hamiltonians: For any orthogonal $M\times M$ matrix $\matO$ we obtain 
\begin{align}\label{eq:mmmatbasis}
\matM_{\mathrm{opt}}[\hat{\rho},\matO\hat{\mathbf{H}},\hat{\mathbf{A}}]=\matO\matR \tilde{\matM}[\hat{\rho},\hat{\mathbf{A}}]\matR^T\matO^T.
\end{align}
Replacing $\hat{\mathbf{H}}$ by $\matO\hat{\mathbf{H}}$ does not affect the optimal measurement observables $\hat{\mathbf{X}}_{\rm opt}$ since $\matO$ can be compensated by the matrix $\matG$ in Eq.~(\ref{eq:optimalX}) for $K=M$. Second, the eigenvalues of $\matM_{\mathrm{opt}}[\hat{\rho},\hat{\mathbf{H}},\hat{\mathbf{A}}]$ are determined by the $M$-dimensional support of $\matR$, which is spanned by $M$ out of the $L$ eigenvectors of $\tilde{\matM}[\hat{\rho},\hat{\mathbf{A}}]$. Since the basis of $\matM_{\mathrm{opt}}[\hat{\rho},\hat{\mathbf{H}},\hat{\mathbf{A}}]$ can be arbitrarily chosen via $\matO$, optimality of the $\matR$ is determined by the spectrum of $\matM_{\mathrm{opt}}[\hat{\rho},\hat{\mathbf{H}},\hat{\mathbf{A}}]$. We consider the parameter encoding optimal if $\matR$ projects onto the subspace corresponding to the $M$ largest eigenvalues of $\tilde{\matM}[\hat{\rho},\hat{\mathbf{A}}]$. For the common case of a shot-noise matrix $\matF_{\mathrm{SN}}[\hat{\mathbf{H}}]$ that is proportional to the $M$-dimensional identity matrix, the same $\matR$ that is optimal for $\matM_{\mathrm{opt}}[\hat{\rho},\hat{\mathbf{H}},\hat{\mathbf{A}}]$ is also optimal for $\matxi^2_{\rm opt}[\hat{\rho},\hat{\mathbf{H}},\hat{\mathbf{A}}]$. Similarly to the optimization over $\hat{\mathbf{X}}$, the phase-imprinting Hamiltonians $\hat{\mathbf{H}}$ must be constrained to physically implementable evolutions.

\textbf{Continuous-variable squeezing matrix.} Families of phase-space operators can be constructed from $\hat{\mathbf{q}}$ by means of a canonical transformation $\matO$ as $\matO\hat{\mathbf{q}}$. Canonical mode transformations are described by $2M\times 2M$ orthogonal symplectic matrices $\matO$ satisfying both $\matO^{-1}=\matO^T$ and $\matO\matOmega\matO^T=\matOmega$ \cite{Ferraro,Wang,Weedbrook}. Notice that the elements of $\matO\hat{\mathbf{q}}$ are in general nonlocal linear combinations of those of $\hat{\mathbf{q}}$, but they follow the same commutation relations.

To discuss the problem of estimating $M$ parameters encoded by the local generators $\hat{\mathbf{H}}=\matR\hat{\mathbf{q}}$, we choose $\matR=\matP_{M}\matO$. Here, the $M\times 2M$ projector $\matP_{M}$ onto canonical basis vectors with even labels picks a single operator (some linear combination of $\hat{x}$ and $\hat{p}$) from each of the local modes in $\matO\hat{\mathbf{q}}$, and $\matO$ is an orthogonal symplectic matrix. This condition ensures that all generators commute: Using the condition $\matR=\matP_{M}\matO$, we find $\tilde{\matC}[\hat{\rho},\hat{\mathbf{H}}]=\frac{1}{2}\matR\matOmega\matR^T=\frac{1}{2}\matP_{M}\matO\matOmega\matO^T\matP_{M}^T=\frac{1}{2}\matP_{M}\matOmega\matP_{M}^T=\matzero_M$, and analogously, $\tilde{\matC}[\hat{\rho},\hat{\mathbf{X}}]=\matzero_M$. Since $\matR\matR^T=\matID_M$, this further implies that the shot-noise limit does not depend on the choice of generators, i.e., $\matF_{\mathrm{SN}}[\hat{\mathbf{H}}]:=\max_{\hat{\rho}_{\rm cl}}\matF_Q[\hat{\rho}_{\rm cl},\hat{\mathbf{H}}]=\matID_{M}$ for all $\hat{\mathbf{H}}=\matR\hat{\mathbf{q}}$.

\textbf{Maximal gain due to mode entanglement.} We first notice that if the squeezing level is identical in all modes, the squeezing matrix becomes proportional to the identity matrix which leaves no room for further optimizations, i.e., all strategies perform equally well. In general, the mode-local state with the diagonal squeezing matrix [$\matU=\matID_M$ in Eq.~(\ref{eq:mmSQZ})] yields an estimation uncertainty of 
\begin{align}\label{eq:localsqzsum}
\mu\mathbf{n}^T\matSigma\mathbf{n}=\mathbf{n}^T\matxi_{\mathrm{opt}}^2[|\Psi_0\rangle,\hat{\mathbf{H}},\hat{\mathbf{q}}]\mathbf{n}=\sum_{k=1}^Mn_k^2e^{-2r_k}.
\end{align}
To identify the corresponding sensitivity limit in the presence of mode entanglement, we change the eigenvectors of the squeezing matrix by applying a passive transformation $\hat{U}_{\matV}$ to the state $|\Psi_0\rangle$. We limit ourselves to passive transformations, since we consider the amount of initial squeezing a fixed resource~\cite{BraunsteinPRA2005}. We show in \supp{4} that passive transformations are sufficient to produce arbitrary basis transformations of the squeezing matrix. Let us denote $\mathbf{n}_1=\mathbf{n}$ and complete it to a basis $\{\mathbf{n}_k\}_{k=1}^M$. Choosing a transformation $\hat{U}_{\matV}$ that achieves $\matxi_{\rm opt}^2[\hat{U}_{\matV}|\Psi_0\rangle,\hat{\mathbf{H}},\hat{\mathbf{q}}]=\sum_{k=1}^Me^{-2r_k}\mathbf{n}_k\mathbf{n}_k^T$, where $r_1\geq \dots\geq r_M$, we obtain
\begin{align}\label{eq:nonlocalsqzsum}
\mu\mathbf{n}^T\matSigma\mathbf{n}=\mathbf{n}^T \matxi_{\rm opt}^2[\hat{U}_{\matV}|\Psi_0\rangle,\hat{\mathbf{H}},\hat{\mathbf{q}}] \mathbf{n}=e^{-2r_1},
\end{align}
which clearly leads to a better precision than~(\ref{eq:localsqzsum}) as long as the squeezing level is not identical in all modes. While Eq.~(\ref{eq:localsqzsum}) makes use of all quadratures and yields the avereage squeezing, weighted by the normalized coefficients $n_k^2$, Eq.~(\ref{eq:nonlocalsqzsum}) maps the maximally squeezed quadrature onto the relevant linear combination of parameters. In other words, we have rotated the state $\hat{U}_{\matV}|\Psi_0\rangle$ such that the smallest eigenvector of $\matxi^2_{\mathrm{opt}}[\hat{U}_{\matV}|\Psi_0\rangle,\hat{\mathbf{H}},\hat{\mathbf{q}}]$ is given by $\mathbf{n}$. Notice that in order to achieve this mapping for a nonlocal $\mathbf{n}$, the state $\hat{U}_{\matV}|\Psi_0\rangle$ becomes mode entangled.

In order to identify the limits of both strategies for a given $\mathbf{n}$, we consider the optimal distribution of a finite total amount of squeezing that minimizes Eq.~(\ref{eq:localsqzsum}) or Eq.~(\ref{eq:nonlocalsqzsum}) for a fixed total average number of particles $N=\sum_{k=1}^M \langle \hat{a}^{\dagger}_k\hat{a}_k\rangle_{|\Psi_0\rangle}=\sum_{k=1}^M\sinh^2 r_k$. 
The constrained minimization of Eq.~(\ref{eq:localsqzsum}) is done with the method of Lagrange multipliers: we write the Lagrange function 
$\mathfrak{L}(\mathbf{x}, \lambda) = \sum_{k=1}^M \tfrac{n_k^2}{x_k} - \lambda \big[\sum_{k=1}^m (\tfrac{x_k}{4}+ \tfrac{1}{4x_k}) - \tfrac{M}{2}-N\big]$,
where $x_k = e^{2r_k}$. The solution of the set of $M+1$ equations $\tfrac{d \mathfrak{L}(\mathbf{x}, \lambda) }{d \lambda}=0$ for $k=1, ..., M$ and 
$\tfrac{d \mathfrak{L}(\mathbf{x}, \lambda) }{d x_k}=0$ gives
$n_k^2 = (\lambda/4) (x_k^2-1)$. Summing over $k$ and imposing $\sum_{k=1}^M n_k^2 = 1$, we find
\begin{align} \label{eq:localsqzsum1}
\frac{e^{4 r_k}-1}{\sum_{k=1}^M (e^{4 r_k} - 1)}= n_k^2,
\end{align}
whose solution gives the optimal squeezing parameters $r_k$.

Clearly, the mode-entangled sensitivity~(\ref{eq:nonlocalsqzsum}) is optimized by concentrating all available squeezing into the initial mode that will be mapped by $\hat{U}_{\matV}$ onto the optimal nonlocal mode, characterized by $\mathbf{n}$, leading to
\begin{align}
(\Delta \theta_{\rm m-ent})^2 = e^{-2 r'},
\end{align}
where $\sinh^2 r ' = \sum_{k=1}^M \sinh^2 r_k$ for the conservation of the total average particle number. 

In the following, let us consider, for simplicity, the estimation of an equally weighted linear combination of all parameters, i.e., $n_k^2=1/M$ for $k=1,\dots,M$. This implies that all the $r_k\equiv r$ are identically chosen and $(\Delta \theta_{\rm m-sep})^2 = e^{-2 r}$, where $r = {\rm arcsinh} \sqrt{N/M}$. The entanglement-enabled noise suppression factor is given by $(\Delta \theta_{\rm m-ent})^2 / (\Delta \theta_{\rm m-sep})^2 = e^{-2 r'}/e^{-2 r}$. In the case $r = 0$ (that also implies $r'=0$), we have  $(\Delta \theta_{\rm m-ent})^2 / (\Delta \theta_{\rm m-sep})^2 = 1$: 
the mode-entangled and mode-separable strategies perform equally well.
When $r \ll 1/\sqrt{M}$ we can approximate $r '  \approx \sqrt{M} r$ (recall that $\sinh^2 r \approx r^2 + O(r^3)$) and obtain 
$(\Delta \theta_{\rm m-ent})^2 / (\Delta \theta_{\rm m-sep})^2 \approx e^{-2 (\sqrt{M}-1) r}$.
When $r \gg1$ (that also implies $r' \gg 1$) we have $e^{-2 r'} \approx M e^{-2r}$.

\begin{acknowledgments}
This work was supported by the LabEx ENS-ICFP: ANR-10-LABX-0010/ANR-10-IDEX-0001-02 PSL* and the European Commission through the QuantERA ERA-NET Cofund in Quantum Technologies project “CEBBEC”. The authors acknowledge financial support from the European Union’s Horizon 2020 research and innovation programme -- Qombs Project, FET Flagship on Quantum Technologies grant no. 820419.
\end{acknowledgments}

\section*{Author contributions}
M.G., A.S., and L.P. contributed to all aspects of this work.

\section*{Data availability}
All relevant data are available from the authors.

\section*{Code availability}
Source codes of the plots are available from the corresponding author upon request.

\section*{Competing interests}
The authors declare no competing interests.

\makeatletter
\renewcommand{\theequation}{S\arabic{equation}}    
\@addtoreset{equation}{section}
\setcounter{equation}{0}
\makeatother

\begin{widetext}
\section*{Supplementary Note 1: The multiparameter method of moments in the central limit}\label{app:CL}
Here, we derive the multiparameter sensitivity matrix for the method of moments in the central limit. Assuming a large number $\mu$ of repeated measurements, the multivariate central limit theorem~\cite{LehmannCasella} ensures that the distribution of  sample mean values $\bar{\mathbf{X}}^{(\mu)}$ approaches a Gaussian multivariate distribution 
\begin{align}\label{eq:multigaussian}
&P(\bar{\mathbf{X}}^{(\mu)}|\vectheta) =\sqrt{\frac{\mu}{(2\pi)^{M}\det\matGamma[\hat{\rho}(\vectheta),\hat{\mathbf{X}}]}}
\exp\left(-\frac{\mu}{2}(\bar{\mathbf{X}}^{(\mu)}-\langle \hat{\mathbf{X}}\rangle_{\hat{\rho}(\vectheta)})^T\matGamma[\hat{\rho}(\vectheta),\hat{\mathbf{X}}]^{-1}(\bar{\mathbf{X}}^{(\mu)}-\langle \hat{\mathbf{X}}\rangle_{\hat{\rho}(\vectheta)})\right) 
\end{align}
with mean $\langle \hat{\mathbf{X}}\rangle_{\hat{\rho}(\vectheta)}$ and covariance matrix $\matGamma[\hat{\rho}(\vectheta),\hat{\mathbf{X}}]/\mu$. The conditions $\langle\hat{X}_k\rangle_{\hat{\rho}(\vectheta)}=\bar{X}_k^{(\mu)}$ identify the multidimensional maximum of Eq.~(\ref{eq:multigaussian}). Hence, the multiparameter method of moments maps to a maximum likelihood estimation of all parameters, which saturates the Cram\'{e}r-Rao bound~\cite{KayBOOK} and asymptotically in the number of measurements $\mu$ leads to the estimator covariance matrix $\matSigma=\matSigma_{\rm mm}$ with
\begin{align}
\matSigma_{\rm mm} = (\mu \matF_{\rm mm}[\hat{\rho}(\vectheta),\hat{\mathbf{X}}])^{-1}.
\end{align}
Here, $\matF_{\rm mm}[\hat{\rho}(\vectheta),\hat{\mathbf{X}}]$
is the Fisher information matrix for the distribution in Eq.~(\ref{eq:multigaussian}) 
(that should not be confused with the Fisher information matrix $\matF[\hat{\rho}(\vectheta),\hat{\mathbf{X}}]$), 
\begin{align}
(\matF_{\rm mm}[\hat{\rho}(\vectheta),\hat{\mathbf{X}}])_{kl}=\sum_{\bar{\mathbf{X}}^{(\mu)}}
p(\bar{\mathbf{X}}^{(\mu)}|\vectheta)
\bigg(\frac{\partial}{\partial \theta_k}\log p(\bar{\mathbf{X}}^{(\mu)}|\vectheta)\bigg)
\bigg(\frac{\partial}{\partial \theta_l}\log p(\bar{\mathbf{X}}^{(\mu)}|\vectheta)\bigg),
\end{align}
the sum running over all possible values of $\bar{\mathbf{X}}^{(\mu)}$ and $p(\bar{\mathbf{X}}^{(\mu)}|\vectheta)$ is the probability to observe the sample mean value $\bar{\mathbf{X}}^{(\mu)}$ given that the parameters take on the values $\vectheta$.
The explicit calculation, see Ref.~\cite{KayBOOK}, 
gives
\begin{align}
(\matF_{\rm mm}[\hat{\rho}(\vectheta),\hat{\mathbf{X}}])_{kl}=\mu\left(\frac{\partial\langle \hat{\mathbf{X}}\rangle_{\hat{\rho}(\vectheta)}}{\partial\theta_k}\right)^T\matGamma[\hat{\rho}(\vectheta),\hat{\mathbf{X}}]^{-1}\left(\frac{\partial\langle \hat{\mathbf{X}}\rangle_{\hat{\rho}(\vectheta)}}{\partial\theta_l}\right)+\frac{1}{2}\mathrm{Tr}\left\{\matGamma[\hat{\rho}(\vectheta),\hat{\mathbf{X}}]^{-1}\left(\frac{\partial}{\partial\theta_k}\matGamma[\hat{\rho}(\vectheta),\hat{\mathbf{X}}]\right)\matGamma[\hat{\rho}(\vectheta),\hat{\mathbf{X}}]^{-1}\left(\frac{\partial}{\partial\theta_l}\matGamma[\hat{\rho}(\vectheta),\hat{\mathbf{X}}]\right)\right\},
\end{align}
where the derivatives of the vectors and matrices are defined element-wise. Since we assume $\mu\gg 1$, the contribution of the first term dominates over the second which thus can be neglected. This yields the result
\begin{align}\label{eq:gaussianfishermatrixS}
\matSigma=(\mu\matM[\hat{\rho}(\vectheta),\hat{\mathbf{X}}])^{-1},
\end{align}
where $\matM[\hat{\rho}(\vectheta),\hat{\mathbf{X}}]$ is the moment matrix.
\end{widetext}

\section*{Supplementary Note 2: Properties of the moment matrix}
The central quantity of interest to characterize (generalized) multiparameter squeezing is the moment matrix, defined by
\begin{align}\label{eq:MDmatrix}
\matM[\hat{\rho}(\vectheta),\hat{\mathbf{X}}]=\matD[\hat{\rho}(\vectheta),\hat{\mathbf{X}}]^T\matGamma[\hat{\rho}(\vectheta),\hat{\mathbf{X}}]^{-1}\matD[\hat{\rho}(\vectheta),\hat{\mathbf{X}}],
\end{align}
where
\begin{align}\label{eq:Dmatrix}
(\matD[\hat{\rho}(\vectheta),\hat{\mathbf{X}}])_{kl}=\frac{\partial\langle\hat{X}_k\rangle_{\hat{\rho}(\vectheta)}}{\partial \theta_l}
\end{align}
and $(\matGamma[\hat{\rho}(\vectheta),\hat{\mathbf{X}}])_{kl}=\langle\hat{X}_k\hat{X}_l\rangle_{\hat{\rho}(\vectheta)}-\langle\hat{X}_k\rangle_{\hat{\rho}(\vectheta)}\langle\hat{X}_l\rangle_{\hat{\rho}(\vectheta)}$ is the covariance matrix with $\langle\hat{X}\rangle_{\hat{\rho}}=\mathrm{Tr}\{\hat{X}\hat{\rho}\}$. An important special case is given by a unitary phase imprinting evolution, when the moment matrix is given by
\begin{align}\label{eq:momentmatrixS}
\matM[\hat{\rho},\hat{\mathbf{H}},\hat{\mathbf{X}}]=\matC[\hat{\rho},\hat{\mathbf{H}},\hat{\mathbf{X}}]^T \, \matGamma[\hat{\rho},\hat{\mathbf{X}}]^{-1} \,
\matC[\hat{\rho},\hat{\mathbf{H}},\hat{\mathbf{X}}],
\end{align}
and $(\matC[\hat{\rho},\hat{\mathbf{H}},\hat{\mathbf{X}}])_{kl}=-i\langle [\hat{X}_k,\hat{H}_l]\rangle_{\hat{\rho}}$ is the commutator matrix. In this section, we prove the main properties of the moment matrix.

\subsection{A matrix-valued Cauchy-Schwarz inequality}\label{app:CS}
We first demonstrate a generalization of the Cauchy-Schwarz inequality to matrices, cf. Refs.~\cite{Chipman1964,Styan1996}. We denote by $\mathrm{Mat}(n,m)$ the space of real-valued $n\times m$ matrices.\\
\textit{Lemma.} Let $\matA\in\mathrm{Mat}(p,n)$, $\matB\in\mathrm{Mat}(p,m)$ and let $\matB^T\matB$ be invertible. Then the following matrix inequality holds:
\begin{align}\label{eq:multiCS}
\matA^T\matA\geq \matA^T\matB(\matB^T\matB)^{-1}\matB^T\matA,
\end{align}
and equality is reached if and only if there is some $\matE\in\mathrm{Mat}(m,n)$ such that
\begin{align}\label{eq:CSsaturation}
\matA=\matB\matE.
\end{align}
\textit{Proof.} For any $\matK\in\mathrm{Mat}(m,n)$, we have that $\matA+\matB\matK\in\mathrm{Mat}(p,n)$ and $(\matA+\matB\matK)^T(\matA+\matB\matK)\geq 0$.  
Inserting $\matK=-(\matB^T\matB)^{-1}\matB^T\matA$ yields 
\begin{align}
&\qquad
\matA^T\matA-\matA^T\matB(\matB^T\matB)^{-1}\matB^T\matA-\matA^T\matB(\matB^T\matB)^{-1}\matB^T\matA\notag\\&\quad
+\matA^T\matB\underbrace{(\matB^T\matB)^{-1}\matB^T\matB(\matB^T\matB)^{-1}}_{(\matB^T\matB)^{-1}}\matB^T\matA\geq 0,\notag
\end{align}
which proves the bound. The saturation condition $\matA+\matB\matK=0$ is satisfied if and only if $\matA=\matPi_{\matB}\matA$, where $\matPi_{\matB}=\matB(\matB^T\matB)^{-1}\matB^T$ is the projector onto the range of $\matB$. This is equivalent to $\matA=\matB\matE$ for some $\matE\in\mathrm{Mat}(m,n)$ and completes the proof.

For $n=m=1$ this inequality reduces to the Cauchy-Schwarz inequality for vectors. The saturation condition~(\ref{eq:CSsaturation}) then yields the well-known requirement that the $p$-dimensional vectors $\matA$ and $\matB$ must be parallel. 

\subsection{General properties}\label{app:convexM}
\textit{Convexity.---}Introducing vectors $\matA$ and $\matB$ with matrix-valued entries $\matA_{\gamma}=\sqrt{p_{\gamma}}\matGamma[\hat{\rho}_{\gamma},\hat{\mathbf{X}}]^{-\frac{1}{2}}\matC[\hat{\rho}_{\gamma},\hat{\mathbf{H}},\hat{\mathbf{X}}]$ and $\matB_{\gamma}=\sqrt{p_{\gamma}}\matGamma[\hat{\rho}_{\gamma},\hat{\mathbf{X}}]^{\frac{1}{2}}$, we obtain $\matA^T\matA=\sum_{\gamma}p_{\gamma}\matC[\hat{\rho}_{\gamma},\hat{\mathbf{H}},\hat{\mathbf{X}}]^T\matGamma[\hat{\rho}_{\gamma},\hat{\mathbf{X}}]^{-1}\matC[\hat{\rho}_{\gamma},\hat{\mathbf{H}},\hat{\mathbf{X}}]$, $\matB^T\matB=\sum_{\gamma}p_{\gamma}\matGamma[\hat{\rho}_{\gamma},\hat{\mathbf{X}}]$, and $\matB^T\matA=\sum_{\gamma}p_{\gamma}\matC[\hat{\rho}_{\gamma},\hat{\mathbf{H}},\hat{\mathbf{X}}]$. From Eq.~(\ref{eq:multiCS}) follows that
\begin{align}
&\quad
\sum_{\gamma}p_{\gamma}\matM[\hat{\rho}_{\gamma},\hat{\mathbf{H}},\hat{\mathbf{X}}]\notag\\&
\geq \matC[\hat{\rho},\hat{\mathbf{H}},\hat{\mathbf{X}}]^T\left(\sum_{\gamma}p_{\gamma}\matGamma[\hat{\rho}_{\gamma},\hat{\mathbf{X}}]\right)^{-1}\matC[\hat{\rho},\hat{\mathbf{H}},\hat{\mathbf{X}}],
\end{align}
where $\matC[\hat{\rho},\hat{\mathbf{H}},\hat{\mathbf{X}}]=\sum_{\gamma}p_{\gamma}\matC[\hat{\rho}_{\gamma},\hat{\mathbf{H}},\hat{\mathbf{X}}]$ for $\hat{\rho}=\sum_{\gamma}p_{\gamma}\hat{\rho}_{\gamma}$. Furthermore, the concavity of the covariance matrix implies that $\left(\sum_{\gamma}p_{\gamma}\matGamma[\hat{\rho}_{\gamma},\hat{\mathbf{X}}]\right)^{-1}\geq \matGamma[\hat{\rho},\hat{\mathbf{X}}]^{-1}$ and we finally obtain the convexity property:
\begin{align}
\matM[\hat{\rho},\hat{\mathbf{H}},\hat{\mathbf{X}}]\leq\sum_{\gamma}p_{\gamma}\matM[\hat{\rho}_{\gamma},\hat{\mathbf{H}},\hat{\mathbf{X}}].
\end{align}

\textit{Orthogonal transformations.---}Let us first note some general transformation properties of the covariance and commutator matrices, respectively. We first introduce a larger family of Hermitian operators $\hat{\mathbf{A}}=(\hat{A}_1,\dots,\hat{A}_L)^T$, such that we can express the elements of $\hat{\mathbf{H}}$ and $\hat{\mathbf{X}}$ as linear combinations of the elements of $\hat{\mathbf{A}}$ (expressed as a column vector):
\begin{align}\label{eq:HXA}
\hat{\mathbf{H}}&=\matR\hat{\mathbf{A}},\qquad\hat{\mathbf{X}}=\matS\hat{\mathbf{A}}.
\end{align}
The following property holds:
\begin{align}\label{eq:GammaXA}
\matGamma[\hat{\rho},\hat{\mathbf{X}}]&=\matS\matGamma[\hat{\rho},\hat{\mathbf{A}}]\matS^T,
\end{align}
where we used that $\mathrm{Cov}(\hat{X}_k,\hat{X}_l)_{\hat{\rho}}=\sum_{ij=1}^L s_{k,i}s_{l,j}\mathrm{Cov}(\hat{A}_i,\hat{A}_j)_{\hat{\rho}}$, due to the bilinearity of the covariance. Analogously,
\begin{align}\label{eq:CXA}
\matC[\hat{\rho},\hat{\mathbf{H}},\hat{\mathbf{X}}]=\matS\tilde{\matC}[\hat{\rho},\hat{\mathbf{A}}]\matR^T
\end{align}
follows from the bilinearity of the commutator, i.e., $-i\langle [\hat{X}_k,\hat{H}_l]\rangle_{\hat{\rho}}=-i\sum_{ij=1}^Ls_{k,i}r_{l,j}\langle [\hat{A}_i,\hat{A}_j]\rangle_{\hat{\rho}}$, and $(\tilde{\matC}[\hat{\rho},\hat{\mathbf{A}}])_{ij}=-i\langle [\hat{A}_i,\hat{A}_j]\rangle_{\hat{\rho}}$. From the definition 
\begin{align}\label{eq:momentmatrixAS}
\tilde{\matM}[\hat{\rho},\hat{\mathbf{A}}]=\tilde{\matC}[\hat{\rho},\hat{\mathbf{A}}]^T\matGamma[\hat{\rho},\hat{\mathbf{A}}]^{-1}\tilde{\matC}[\hat{\rho},\hat{\mathbf{A}}],
\end{align}
and the transformation properties~(\ref{eq:GammaXA}) and~(\ref{eq:CXA}) follows for $\matS=\matR=\matO$, where $\matO$ is an orthogonal matrix, that
\begin{align}\label{eq:transMA}
\tilde{\matM}[\hat{\rho},\matO\hat{\mathbf{A}}]=\matO\tilde{\matM}[\hat{\rho},\hat{\mathbf{A}}]\matO^T.
\end{align}
The matrix $\matO$ can be chosen to diagonalize the moment matrix.

\subsection{Maximizing the moment matrix}\label{app:maxX}
Here we maximize the moment matrix~(\ref{eq:momentmatrixS}) over the measurement observables $\hat{\mathbf{X}}$ as a function of a family of measurable operators $\hat{\mathbf{A}}=(\hat{A}_1,\dots,\hat{A}_L)^T$. Let $\hat{\mathbf{X}}=(\hat{X}_1,\dots,\hat{X}_K)^T$ and $\hat{\mathbf{H}}=(\hat{H}_1,\dots,\hat{H}_M)^T$, i.e., $\matR\in\mathrm{Mat}(M,L)$ and $\matS\in\mathrm{Mat}(K,L)$.

Inserting Eqs.~(\ref{eq:GammaXA}) and~(\ref{eq:CXA}) into Eq.~(\ref{eq:momentmatrixS}) yields the following expression for the moment matrix:
\begin{align}\notag
\matM[\hat{\rho},\hat{\mathbf{H}},\hat{\mathbf{X}}]=\matR\tilde{\matC}[\hat{\rho},\hat{\mathbf{A}}]^T\matS^T\left(\matS\matGamma[\hat{\rho},\hat{\mathbf{A}}]\matS^T\right)^{-1}\matS\tilde{\matC}[\hat{\rho},\hat{\mathbf{A}}]\matR^T.
\end{align}

We now apply the inequality~(\ref{eq:multiCS}) with $\matA=\matGamma[\hat{\rho},\hat{\mathbf{A}}]^{-\frac{1}{2}}\tilde{\matC}[\hat{\rho},\hat{\mathbf{A}}]\matR^T$ and $\matB=\matGamma[\hat{\rho},\hat{\mathbf{A}}]^{\frac{1}{2}}\matS^T$, leading to
\begin{align}\label{eq:momentsboundapp}
\matM[\hat{\rho},\hat{\mathbf{H}},\hat{\mathbf{X}}]\leq \matR\tilde{\matM}[\hat{\rho},\hat{\mathbf{A}}]\matR^T,
\end{align}
with $\tilde{\matM}[\hat{\rho},\hat{\mathbf{A}}]$ defined in Eq.~(\ref{eq:momentmatrixAS}).

Inequality~(\ref{eq:momentsboundapp}) provides an upper bound on the moment-based multiparameter sensitivity for any choice of the observables $\hat{\mathbf{X}}$. The maximal sensitivity is reached when the inequality is saturated. The saturation condition~(\ref{eq:CSsaturation}) is fulfilled when 
\begin{align}\label{eq:optXapp}
\matG\matS=\matR\tilde{\matC}[\hat{\rho},\hat{\mathbf{A}}]^T\matGamma[\hat{\rho},\hat{\mathbf{A}}]^{-1},
\end{align}
for some $\matG\in\mathrm{Mat}(M,K)$, which corresponds to $\matG=\matE^T$ in Eq.~(\ref{eq:CSsaturation}). Recall that the matrix $\matS$ determines the measurement operators $\hat{\mathbf{X}}$ via Eq.~(\ref{eq:HXA}). The freedom provided by the matrix $\matG$ can be used to rearrange and normalize the measurement observables. When we have as many measurement observables as there are parameters $K=M$, we can choose $\matG=\matT^{-1}$, leading to the expression
\begin{align}\label{eq:optimalXS}
\hat{\mathbf{X}}_{\mathrm{opt}}=\matT\matR\tilde{\matC}[\hat{\rho},\hat{\mathbf{A}}]^T\matGamma[\hat{\rho},\hat{\mathbf{A}}]^{-1}\hat{\mathbf{A}},
\end{align}
where $\matT$ is an arbitrary invertible $M\times M$ matrix. By demonstrating the saturability of the lower bound~(\ref{eq:momentsboundapp}), we have solved the maximization problem of the moment matrix over all measurement operators $\hat{\mathbf{X}}$ from the accessible set $\hat{\mathbf{A}}$ for fixed $\hat{\mathbf{H}}$. In practice, saturation can be achieved when all elements of $\hat{\mathbf{X}}_{\mathrm{opt}}$ can be measured simultaneously. 

\subsection{Lower bound on the classical Fisher matrix}\label{app:lowerbound}
Here, we demonstrate a general result that implies 
\begin{align}\label{eq:boundMFS}
\max_{\hat{\mathbf{X}}\in\mathrm{span}(\hat{\bs{\Pi}})}\matM[\hat{\rho}(\vectheta),\hat{\mathbf{H}},\hat{\mathbf{X}}]=\matF[\hat{\rho}(\vectheta),\hat{\mathbf{X}}]
\end{align}
for the special case of unitary evolution. We denote the spectral decomposition of the observables as $\hat{X}_l=\sum_{k}x_l(k)\hat{\Pi}_{k}$, where the $\hat{\Pi}_{k}$ are the projectors onto a common eigenbasis of all $\hat{X}_l$. The quantum mechanical expectation values are given by $\langle\hat{X}_l\rangle_{\hat{\rho}(\vectheta)}=\sum_{k}x_l(k)p(k|\vectheta)$ with $p(k|\vectheta)=\mathrm{Tr}\{\hat{\rho}(\vectheta)\hat{\Pi}_{k}\}$.

Let us now consider the matrix bound~(\ref{eq:multiCS}) with matrices $\matA_{kl}=\sqrt{p(k|\vectheta)}\left(\frac{\partial}{\partial \theta_l}\log p(k|\vectheta)\right)$ and $\matB_{kl}=\sqrt{p(k|\vectheta)}\left(x_l(k)-\langle\hat{X}_l\rangle_{\hat{\rho}(\vectheta)}\right)$. We obtain
\begin{align}
\matA^T\matA&=\matF[\hat{\rho}(\vectheta),\hat{\mathbf{X}}],\notag\\
\matB^T\matB&=\matGamma[\hat{\rho}(\vectheta),\hat{\mathbf{X}}],\notag\\
\matB^T\matA&=\matD[\hat{\rho}(\vectheta),\hat{\mathbf{X}}],
\end{align}
and~(\ref{eq:multiCS}) implies that
\begin{align}\label{eq:lowerbound}
\matM[\hat{\rho}(\vectheta),\hat{\mathbf{H}},\hat{\mathbf{X}}]\leq\matF[\hat{\rho}(\vectheta),\hat{\mathbf{X}}],
\end{align}
for any $\hat{\rho}(\vectheta)$ and any $\hat{\mathbf{X}}$.

\textit{Saturation condition.---}Let us now discuss the conditions for the saturation of~(\ref{eq:lowerbound}). A straightforward solution to the saturation condition~(\ref{eq:CSsaturation}) is obtained by requiring that $\matA=\matB$, which is achieved by the choice
\begin{align}\label{eq:optspec}
x_l(k)=\frac{\partial}{\partial \theta_l}\log p(k|\vectheta)+\langle\hat{X}_l\rangle_{\hat{\rho}(\vectheta)},
\end{align}
for all $k$ and $l$. This means that a measurement of the observables 
\begin{align}
\hat{X}_l=\sum_{k}\left(\frac{\partial}{\partial \theta_l}\log p(k|\vectheta)+\langle\hat{X}_l\rangle_{\hat{\rho}(\vectheta)}\right)\hat{\Pi}_{k}.
\end{align}
leads to saturation of the bound~(\ref{eq:lowerbound}), i.e., the moment matrix associated with these observables coincides with the Fisher matrix generated by the projectors in their common eigenbasis. Notice that transformations of the type $\hat{X}_l\rightarrow\alpha\hat{X}_l+\beta\hat{\mathbb{I}}_l$ with arbitrary $\alpha,\beta\in\mathbb{R}$ do not alter the moment matrix.

\textit{Saturating the bound by measuring projectors.---}Alternatively, we may consider the special case where the projectors $\hat{\Pi}_{k}$ themselves are the measurement observables and we estimate the parameters $\vectheta$ from the average values of $\hat{\mathbf{X}}=\hat{\bs{\Pi}}$. Recall that the moment matrix~(\ref{eq:MDmatrix}) depends on the inverse of the covariance matrix of the measured observables. Singularities of the covariance matrix may arise either due to redundant information from the measurement of too many projectors that span a complete basis, or from projectors that are orthogonal to the state and lead to $p(k|\vectheta)=\langle \hat{\Pi}_k\rangle_{\rho(\vectheta)}=0$. This can be avoided by effectively limiting the set of measured observables $\hat{\mathbf{X}}=\{\hat{\Pi}_k\}_{k=1}^{d-1}$ to a subset of $d-1$ projectors such that $\sum_{k=1}^{d-1}p(k|\vectheta)=1-p(d|\vectheta)<1$. We obtain from Eq.~(\ref{eq:Dmatrix}) that $\matD[\hat{\rho}(\vectheta),\hat{\mathbf{X}}]_{kl}=\frac{\partial p(k|\vectheta)}{\partial \theta_l}$ and~\cite{GessnerPRA2019}
\begin{align}
\matGamma[\hat{\rho}(\vectheta),\hat{\mathbf{X}}]&=\matP(\vectheta)-\mathbf{p}(\vectheta)\mathbf{p}(\vectheta)^T,
\end{align}
where $\matP(\vectheta)=\mathrm{diag}(p(1|\vectheta),\dots,p(d-1|\vectheta))$ and $\mathbf{p}(\vectheta)=(p(1|\vectheta),\dots,p(d-1|\vectheta))^T$. Using $\matGamma[\hat{\rho}(\vectheta),\hat{\mathbf{X}}]^{-1}=\matP(\vectheta)^{-1}+\frac{1}{p(d|\vectheta)}\mathbf{e}\mathbf{e}^T$ with $\mathbf{e}=(1,\dots,1)^T$ we obtain [for unitary evolutions, this coincides with $\matM[\hat{\rho}(\vectheta),\hat{\mathbf{H}},\hat{\mathbf{X}}]$ defined in Eq.~(\ref{eq:momentmatrixS})]
\begin{align}
\matM[\hat{\rho}(\vectheta),\hat{\mathbf{X}}]&=\matD[\hat{\rho}(\vectheta),\hat{\mathbf{X}}]^T\matP(\vectheta)^{-1}\matD[\hat{\rho}(\vectheta),\hat{\mathbf{X}}]\notag\\&\quad+\frac{1}{p(d|\vectheta)}\left(\matD[\hat{\rho}(\vectheta),\hat{\mathbf{X}}]^T\mathbf{e}\right)\left(\matD[\hat{\rho}(\vectheta),\hat{\mathbf{X}}]^T\mathbf{e}\right)^T.
\end{align}
Using
\begin{align}
&\quad\left(\matD[\hat{\rho}(\vectheta),\hat{\mathbf{X}}]^T\matP(\vectheta)^{-1}\matD[\hat{\rho}(\vectheta),\hat{\mathbf{X}}\right)_{kl}\notag\\&=\sum_{m=1}^{d-1}\frac{1}{p(m|\vectheta)}\left(\frac{\partial p(m|\vectheta)}{\partial \theta_k}\right)\left(\frac{\partial p(m|\vectheta)}{\partial \theta_l}\right)
\end{align}
and
\begin{align}
\left(\matD[\hat{\rho}(\vectheta),\hat{\mathbf{X}}]^T\mathbf{e}\right)_{k}=\sum_{l=1}^{d-1}\frac{\partial p(l|\vectheta)}{\partial \theta_k}=-\frac{\partial p(d|\vectheta)}{\partial \theta_k},
\end{align}
we obtain
\begin{align}\label{eq:Mproj}
\matM[\hat{\rho}(\vectheta),\hat{\mathbf{X}}]_{kl}=\sum_{m=1}^{d}p(m|\vectheta)\left(\frac{\partial }{\partial \theta_k}\log p(m|\vectheta)\right)\left(\frac{\partial }{\partial \theta_l}\log p(m|\vectheta)\right),
\end{align}
which is the Fisher matrix $\matF[\hat{\rho}(\vectheta),\hat{\mathbf{X}}]$. 

We end this section with two remarks: First, we note that we obtain a sensitivity of the form~(\ref{eq:Mproj}) even if the projectors are not of rank one, i.e., when we coarse-grain over measurement outcomes. Second, the result holds for arbitrary choices of the projectors even in the case of non-commuting generators. For pure probe states, necessary and sufficient conditions given in Ref.~\cite{PezzePRL2017} reveal whether a chosen set of projectors also saturates the quantum Fisher matrix.

\subsection{Lower bound on the quantum Fisher matrix}\label{app:MleqF}
Here, we generalize and demonstrate
\begin{align}\label{eq:mHXleqFQ}
\matM[\hat{\rho},\hat{\mathbf{H}},\hat{\mathbf{X}}]\leq \matF_Q[\hat{\rho},\hat{\mathbf{H}}].
\end{align}
Let us first demonstrate that the lower bound
\begin{align}\label{eq:MAFQA}
\tilde{\matM}[\hat{\rho},\hat{\mathbf{A}}] \leq \matF_Q[\hat{\rho},\hat{\mathbf{A}}],
\end{align}
holds for arbitrary families of operators $\hat{\mathbf{A}}$. Here, $(\matF_Q[\hat{\rho},\hat{\mathbf{A}}])_{ij}=\mathrm{Tr}\{\hat{\rho}(\hat{\mathcal{L}}_i\hat{\mathcal{L}}_j+\hat{\mathcal{L}}_j\hat{\mathcal{L}}_i)/2\}$ are the elements of the quantum Fisher matrix with symmetric logarithmic derivatives defined as the solution to $-i[\hat{A}_j,\hat{\rho}]=(\hat{\mathcal{L}}_{j}\hat{\rho}+\hat{\rho}\hat{\mathcal{L}}_{j})/2$ \cite{HelstromBOOK}. 

To prove inequality~(\ref{eq:MAFQA}), note that for an arbitrary $\mathbf{n}=(n_1,\dots,n_L)^T\in\mathbb{R}^L$, the equality $\mathbf{n}^T\tilde{\matM}[\hat{\rho},\hat{\mathbf{A}}]\mathbf{n}=|\langle[\hat{H}_{\mathbf{m}_{\mathrm{opt}}},\hat{H}_{\mathbf{n}}]\rangle_{\hat{\rho}}|^2/(\Delta \hat{H}_{\mathbf{m}_{\mathrm{opt}}})_{\hat{\rho}}^2$ is achieved for an optimally chosen $\hat{H}_{\mathbf{m}}=\mathbf{m}^T\hat{\mathbf{A}}$, where $\mathbf{m}_{\mathrm{opt}}=\alpha\matGamma[\hat{\rho},\hat{\mathbf{A}}]^{-1}\tilde{\matC}[\hat{\rho},\hat{\mathbf{A}}]\mathbf{n}$ with some normalization constant $\alpha$ \cite{GessnerPRL2019}. Furthermore, the inequality $|\langle[\hat{H}_{\mathbf{m}},\hat{H}_{\mathbf{n}}]\rangle_{\hat{\rho}}|^2/(\Delta \hat{H}_{\mathbf{m}})_{\hat{\rho}}^2\leq \mathbf{n}^T\matF_Q[\hat{\rho},\hat{\mathbf{A}}]\mathbf{n}$ holds for all $\mathbf{n},\mathbf{m}\in\mathbb{R}^L$ and saturation is achieved for $\hat{H}_{\mathbf{m}}=\hat{L}_{\mathbf{n}}$ \cite{GessnerPRL2019}, where $\hat{L}_{\mathbf{n}}$ is defined as the solution to the equation
\begin{align}\label{eq:SLDn}
-i[\hat{H}_{\mathbf{n}},\hat{\rho}]=(\hat{L}_{\mathbf{n}}\hat{\rho}+\hat{\rho}\hat{L}_{\mathbf{n}})/2.
\end{align}
We thus obtain $\mathbf{n}^T\tilde{\matM}[\hat{\rho},\hat{\mathbf{A}}]\mathbf{n}\leq \mathbf{n}^T\matF_Q[\hat{\rho},\hat{\mathbf{A}}]\mathbf{n}$ for all $\mathbf{n}$, demonstrating the statement~(\ref{eq:MAFQA}). 

Let us now turn our attention to the saturation condition. The above derivation shows that the equality $\tilde{\matM}[\hat{\rho},\hat{\mathbf{A}}]=\matF_Q[\hat{\rho},\hat{\mathbf{A}}]$ can be achieved if for all $\mathbf{n}$, the $\hat{L}_{\mathbf{n}}$ can be expressed as linear combinations of the elements of $\hat{\mathbf{A}}$. Using the linearity of the condition~(\ref{eq:SLDn}), we find that $L_{\mathbf{n}}=\mathbf{n}^T\hat{\boldsymbol{\mathcal{L}}}$, where $\hat{\boldsymbol{\mathcal{L}}}=(\hat{\mathcal{L}}_1,\dots,\hat{\mathcal{L}}_L)^T$. A sufficient saturation condition is therefore that for each $\hat{A}_j$, also the corresponding $\hat{\mathcal{L}}_j$ is an element of $\hat{\mathbf{A}}$.

Equation~(\ref{eq:mHXleqFQ}) now follows from~(\ref{eq:MAFQA}) by using $\matF_Q[\hat{\rho},\hat{\mathbf{H}}]=\matR\matF_Q[\hat{\rho},\hat{\mathbf{A}}]\matR^T$ \cite{GessnerPRL2018} and Eq.~(\ref{eq:momentsboundapp}). The maximum can be attained if the optimal set of observables~(\ref{eq:optXapp}) can be measured simultaneously.

\section*{Supplementary Note 3: Spin squeezing matrix}\label{app:optmeasSpin}
\setcounter{subsection}{0}
This section discusses properties of the squeezing matrix in the case of discrete variables. Linear parameter-encoding Hamiltonians $\hat{\mathbf{H}}$ and measurement observables $\hat{\mathbf{X}}$ of a collective spin system can be expressed in terms of the $3M$ angular momentum operators 
$\hat{\mathbf{J}}=(\hat{\mathbf{J}}_1^T, ..., \hat{\mathbf{J}}_M^T)^T$ with $\hat{\mathbf{J}}_k=(\hat{J}_{x,k},\hat{J}_{y,k},\hat{J}_{z,k})^T$. 
\subsection{Accessible operators for collective spin systems}\label{app:orthogonalobservables}
Let us first consider the full vector $\hat{\mathbf{J}}$ as family of accessible operators. In this case we obtain the commutator matrix
\begin{align}\label{eq:ClclJ}
\tilde{\matC}[\hat{\rho},\hat{\mathbf{J}}]=\bigoplus_{k=1}^M\begin{pmatrix}0 & \langle\hat{J}_{z,k}\rangle_{\hat{\rho}} & -\langle\hat{J}_{y,k}\rangle_{\hat{\rho}}\\
-\langle\hat{J}_{z,k}\rangle_{\hat{\rho}} & 0 & \langle\hat{J}_{x,k}\rangle_{\hat{\rho}}\\
\langle\hat{J}_{y,k}\rangle_{\hat{\rho}} & -\langle\hat{J}_{x,k}\rangle_{\hat{\rho}} &0\end{pmatrix}.
\end{align}
As a skew-symmetric matrix with odd dimension $3M$, the matrix~(\ref{eq:ClclJ}) is singular due to Jacobi's theorem. The singularity can be avoided by restricting the operator basis to the relevant subset. We define a local mean spin direction $\mathbf{n}_{0,k}=\langle\hat{\mathbf{J}}_k\rangle_{\hat{\rho}}/|\langle\hat{\mathbf{J}}_k\rangle_{\hat{\rho}}|$, which can be extended by two orthogonal vectors $\mathbf{n}_{\perp_1,k}$ and $\mathbf{n}_{\perp_2,k}$ to a complete basis. We can choose any three orthogonal spin operators as a basis to represent linear operators $\hat{H}_k$ and $\hat{X}_k$. Locally rearranging the basis to align the mean field direction $\mathbf{n}_{0,k}$ onto the $z$-direction leads to the commutator matrix
\begin{align}\label{eq:Comega}
\tilde{\matC}[\hat{\rho},\hat{\mathbf{J}}]=\bigoplus_{k=1}^M\begin{pmatrix}0 & \langle\hat{J}_{z,k}\rangle_{\hat{\rho}} & 0\\
-\langle\hat{J}_{z,k}\rangle_{\hat{\rho}} & 0 & 0\\
0 & 0 &0\end{pmatrix}.
\end{align}
As a consequence, the rows and columns of $\tilde{\matM}[\hat{\rho},\hat{\mathbf{J}}]$ [see Eq.~(\ref{eq:momentmatrixAS})] belonging to $\hat{J}_{z,k}$ remain zero and it suffices to restrict to the two-dimensional subspace spanned by $\mathbf{n}_{\perp_1,k}$ and $\mathbf{n}_{\perp_2,k}$. In this subspace, the matrix $\tilde{\matC}[\hat{\rho},\hat{\mathbf{J}}_{\perp}]=\bigoplus_{k=1}^M|\langle\hat{\mathbf{J}}_k\rangle_{\hat{\rho}}|\boldsymbol{\omega}$ with $\boldsymbol{\omega}=\left(\begin{smallmatrix} 0 & 1\\-1 & 0 \end{smallmatrix}\right)$ is always invertible and reflects the symplectic form of canonical transformations, that prominently appears in the description of continuous-variable systems~\cite{Ferraro,Wang,Weedbrook}, locally rescaled by the spin length $\langle\hat{J}_{z,k}\rangle_{\hat{\rho}}=|\langle\hat{\mathbf{J}}_k\rangle_{\hat{\rho}}|$. For simplicity, we henceforth choose a basis described by $\{\mathbf{e}_x,\mathbf{e}_y,\mathbf{e}_z\}=\{\mathbf{n}_{\perp_1,k},\mathbf{n}_{\perp_2,k},\mathbf{n}_{0,k}\}$, such that $\hat{\mathbf{J}}_{\perp}=(\hat{J}_{x,1},\hat{J}_{y,1},\dots,\hat{J}_{x,M},\hat{J}_{y,M})^T$. The operator basis $\hat{\mathbf{J}}_{\perp}$ is sufficient for our purposes since measurements and evolutions that involve the mean spin direction $\mathbf{n}_{0,k}$ are suboptimal for metrology.

\subsection{Multiparameter shot-noise limit}\label{app:SNL}
Here we identify the shot-noise limit for the quantum Fisher matrix generated by the set of accessible operators  $\hat{\mathbf{J}}_{\perp}$. First, we consider the larger set $\hat{\mathbf{J}}$ and we obtain $\sup_{\hat{\rho}_{\mathrm{p-sep}}}\matF_Q[\hat{\rho}_{\mathrm{p-sep}},\hat{\mathbf{J}}]=\sup_{|\Psi^{(1)}\rangle\otimes\cdots\otimes|\Psi^{(N)}\rangle}4\matGamma[|\Psi^{(1)}\rangle\otimes\cdots\otimes|\Psi^{(N)}\rangle,\hat{\mathbf{J}}]$ due to the convexity of the quantum Fisher matrix~\cite{GessnerPRL2018}. Let us now express the vector $\hat{\mathbf{J}}$ as a sum of single-particle vectors $\hat{\mathbf{J}}=\sum_{i=1}^N\hat{\mathbf{J}}^{(i)}$, where $\hat{\mathbf{J}}^{(i)}=(\hat{j}^{(i)}_{x,1},\hat{j}^{(i)}_{y,1},\hat{j}^{(i)}_{z,1},\hat{j}^{(i)}_{x,2},\hat{j}^{(i)}_{y,2},\hat{j}^{(i)}_{z,2},\dots,\hat{j}^{(i)}_{x,M},\hat{j}^{(i)}_{y,M},\hat{j}^{(i)}_{z,M})^T$. The operators $\hat{j}^{(i)}_{\alpha,k}=\frac{1}{2}\hat{\Pi}^{(i)}_k\hat{\sigma}^{(i)}_{\alpha,k}\hat{\Pi}^{(i)}_k$ contain the projectors $\hat{\Pi}^{(i)}_k$ of particle $i$ onto the subspace of mode $k$, such that $\hat{\Pi}^{(i)}_k\hat{\Pi}^{(j)}_l=\delta_{kl}\delta_{ij}\hat{\Pi}^{(i)}_k$. We may expand the single-particle states as $|\Psi^{(i)}\rangle=\sum_{k=1}^M\sqrt{p^{(i)}_k}|\Psi^{(i)}_k\rangle$, with orthonormal local states $\hat{\Pi}^{(i)}_l|\Psi^{(i)}_k\rangle=\delta_{kl}|\Psi^{(i)}_k\rangle$ and $p^{(i)}_k$ denotes the probability for particle $i$ to be in mode $k$ with $\sum_{k=1}^Mp^{(i)}_k=1$. Using $\matGamma[|\Psi^{(1)}\rangle\otimes\cdots\otimes|\Psi^{(N)}\rangle,\hat{\mathbf{J}}]=\sum_{i=1}^N\matGamma[|\Psi^{(i)}\rangle,\hat{\mathbf{J}}^{(i)}]$ and discarding the first moments \cite{GessnerPRL2018}, we obtain
\begin{align}\label{eq:spGamma}
&\quad
\matGamma[|\Psi^{(1)}\rangle\otimes\cdots\otimes|\Psi^{(N)}\rangle,\hat{\mathbf{J}}]\notag\\&
\leq \sum_{i=1}^N\begin{pmatrix}p^{(i)}_1\tilde{\matGamma}[|\Psi_1^{(i)}\rangle,\hat{\mathbf{j}}_1^{(i)}] & \matzero & \cdots & \matzero\\
\vdots & \ddots & & \vdots\\
\matzero & \cdots & \matzero & p^{(i)}_M\tilde{\matGamma}[|\Psi_M^{(i)}\rangle,\hat{\mathbf{j}}_M^{(i)}] 
\end{pmatrix},
\end{align}
where $\hat{\mathbf{j}}_k^{(i)}=(\hat{j}^{(i)}_{x,k},\hat{j}^{(i)}_{y,k},\hat{j}^{(i)}_{z,k})^T$, the $\tilde{\matGamma}[|\Psi_k^{(i)}\rangle,\hat{\mathbf{j}}_k^{(i)}]$ are $3\times 3$ matrices with elements $(\tilde{\matGamma}[|\Psi_k^{(i)}\rangle,\hat{\mathbf{j}}_k^{(i)}])_{\alpha\beta}=\frac{1}{2}\langle \hat{j}_{\alpha,k}^{(i)}\hat{j}_{\beta,k}^{(i)}+\hat{j}_{\beta,k}^{(i)}\hat{j}_{\alpha,k}^{(i)}\rangle_{|\Psi_k^{(i)}\rangle}=\frac{1}{8}\langle \hat{\sigma}_{\alpha,k}^{(i)}\hat{\sigma}_{\beta,k}^{(i)}+\hat{\sigma}_{\beta,k}^{(i)}\hat{\sigma}_{\alpha,k}^{(i)}\rangle_{|\Psi_k^{(i)}\rangle}$, and $\matzero$ is the $3\times 3$ zero matrix. Using the anticommutativity property of the Pauli matrices we obtain that $\tilde{\matGamma}[|\Psi_k^{(i)}\rangle,\hat{\mathbf{j}}_k^{(i)}]=\frac{1}{4}\matID$ for all $k$, $i$ and arbitrary $|\Psi_k^{(i)}\rangle$, where $\matID$ is the $3\times 3$ identity matrix. The upper bound $\matF_Q[\hat{\rho}_{\mathrm{p-sep}},\hat{\mathbf{J}}]\leq \mathrm{diag}(N_1\matID,\dots,N_M\matID)$ is obtained by inserting this back into Eq.~(\ref{eq:spGamma}), and using that $\sum_{i=1}^Np^{(i)}_k=N_k$ is the average number of particles in mode $k$. This upper bound cannot be saturated since not all first moments can be zero simultaneously for a pure single-qubit state. However, by restricting the set of accessible operators to the two directions orthogonal to the mean-spin direction $\mathbf{n}_{0,k}$, we obtain an analogous result with $2\times 2$ instead of $3\times 3$ blocks, i.e., $\matF_Q[\hat{\rho}_{\mathrm{p-sep}},\hat{\mathbf{J}}_{\perp}]\leq\mathrm{diag}(N_1,N_1,\dots,N_M,N_M)$. This bound is saturated by single-qubit states $|\Psi^{(i)}_k\rangle$ that are polarized along $\mathbf{n}_{0,k}$.

\subsection{Local squeezing}\label{app:lclsqz}
In the case of an uncorrelated product of single-mode squeezed states $\hat{\rho}_{\mathrm{loc}}=\hat{\rho}_1\otimes\cdots\otimes\hat{\rho}_M$, the moment matrix attains a block-diagonal form, $\tilde{\matM}[\hat{\rho}_{\rm loc},\hat{\mathbf{J}}_{\perp}]=\bigoplus_{k=1}^M\tilde{\matM}[\hat{\rho}_k,\hat{\mathbf{J}}_{\perp,k}]$. Furthermore, $\matF_{\mathrm{SN}}[\hat{\mathbf{H}}]=\matR\matF_{\mathrm{SN}}[\hat{\mathbf{J}}_{\perp}]\matR^T=\mathrm{diag}(N_1,\dots,N_M)$ implies that the squeezing matrix, optimized over all measurement observables, reads
\begin{align}\label{eq:xioptS}
\matxi_{\mathrm{opt}}^{2}[\hat{\rho}_{\rm loc},\hat{\mathbf{H}},\hat{\mathbf{J}}_{\perp}]&=\min_{\hat{\mathbf{X}}\in\mathrm{span}(\hat{\mathbf{A}})}\matxi^{2}[\hat{\rho}_{\rm loc},\hat{\mathbf{H}},\hat{\mathbf{X}}]\notag\\
&=\matF_{\mathrm{SN}}[\hat{\mathbf{H}}]^{\frac{1}{2}}\matR\tilde{\matM}[\hat{\rho}_{\rm loc},\hat{\mathbf{J}}_{\perp}]^{-1}\matR^T\matF_{\mathrm{SN}}[\hat{\mathbf{H}}]^{\frac{1}{2}}\notag\\
&=\matR\bigoplus_{k=1}^MN_k\tilde{\matM}[\hat{\rho}_k,\hat{\mathbf{J}}_{\perp,k}]^{-1}\matR^T.
\end{align}

\subsubsection{Optimizing the phase-imprinting Hamiltonians}
An optimal choice for $\matR$ is provided when the eigenvalues of the $M\times M$ matrix $\matxi^2_{\mathrm{opt}}$ correspond to the $M$ smallest eigenvalues of the $2M\times 2M$ matrix $\bigoplus_{k=1}^MN_k\tilde{\matM}[\hat{\rho}_k,\hat{\mathbf{J}}_{\perp,k}]^{-1}$ (see also the discussion in the Methods section). Let us therefore consider the eigenvalues of this matrix. Each of the $2\times 2$ blocks can be written as 
\begin{align}\label{eq:mblocks}
N_k\tilde{\matM}[\hat{\rho}_k,\hat{\mathbf{J}}_{\perp,k}]^{-1}=\frac{N_k}{\langle\hat{J}_{z,k}\rangle_{\hat{\rho}_k}^{2}}\matomega\matGamma[\hat{\rho}_k,\hat{\mathbf{J}}_{\perp,k}]\matomega^T,
\end{align}
where we used that $\matC[\hat{\rho}_k,\hat{\mathbf{J}}_{\perp,k}]=\langle\hat{J}_{z,k}\rangle_{\hat{\rho}_k}\matomega$ with $\matomega=\left(\begin{smallmatrix} 0 & 1\\-1 & 0 \end{smallmatrix}\right)$ [see Eq.~(\ref{eq:Comega})]. The eigenvalues
\begin{align}
\lambda_{-,k}&=N_k\min_{\substack{\mathbf{r}_k\\|\mathbf{r}_k|^2=1}}\mathbf{r}_k^T\tilde{\matM}[\hat{\rho}_k,\hat{\mathbf{J}}_{\perp,k}]^{-1}\mathbf{r}_k=N_k\lambda_{\max}(\tilde{\matM}[\hat{\rho}_k,\hat{\mathbf{J}}_{\perp,k}])^{-1},\notag\\
\lambda_{+,k}&=N_k\max_{\substack{\mathbf{r}_k\\|\mathbf{r}_k|^2=1}}\mathbf{r}_k^T\tilde{\matM}[\hat{\rho}_k,\hat{\mathbf{J}}_{\perp,k}]^{-1}\mathbf{r}_k=N_k\lambda_{\min}(\tilde{\matM}[\hat{\rho}_k,\hat{\mathbf{J}}_{\perp,k}])^{-1}\notag
\end{align}
correspond to a squeezed and an anti-squeezed variance (renormalized by the mean spin length), respectively. Indeed, $\lambda_{-,k}$ can be identified as the single-mode spin-squeezing coefficient of mode $k$ \cite{Wineland, Wineland2,RMP,MaPHYSREP2011}, optimized over all local measurements and evolutions \cite{GessnerPRL2019}, $\lambda_{-,k}=\xi_{\min}^{2}[\hat{\rho}_{k},\hat{\mathbf{J}}_{\perp,k}]=N_k\min_{\mathbf{r}_k,\mathbf{s}_k}|\langle[\hat{J}_{\mathbf{s}_k,k},\hat{J}_{\mathbf{r}_k,k}]\rangle_{\hat{\rho}_k}|^{-2}(\Delta \hat{J}_{\mathbf{s}_k})^{2}_{\hat{\rho}_k}$. The uncertainty relation $(\Delta\hat{J}_{x,k})_{\hat{\rho}_k}(\Delta\hat{J}_{y,k})_{\hat{\rho}_k}\geq |\langle\hat{J}_{z,k}\rangle_{\hat{\rho}_k}|/2$ excludes that both directions of the same mode can be simultaneously squeezed, i.e., $\xi_{\min}^2[\hat{\rho}_k,\hat{\mathbf{J}}_{\perp,k}]<1$ implies that $\lambda_{+,k}>1$: We have $\mathrm{Tr}\{\matGamma[\hat{\rho}_k,\hat{\mathbf{J}}_{\perp,k}]\}=(\Delta\hat{J}_{x,k})_{\hat{\rho}_k}^2+(\Delta\hat{J}_{y,k})_{\hat{\rho}_k}^2\geq 2(\Delta\hat{J}_{x,k})_{\hat{\rho}_k}(\Delta\hat{J}_{y,k})_{\hat{\rho}_k}\geq |\langle\hat{J}_{z,k}\rangle_{\hat{\rho}_k}|$, which leads to $\lambda_{+,k}+\lambda_{-,k}=N_k\mathrm{Tr}\{\tilde{\matM}[\hat{\rho}_k,\hat{\mathbf{J}}_{\perp,k}]^{-1}\}=N_k\langle\hat{J}_{z,k}\rangle^{-2}\mathrm{Tr}\{\matGamma[\hat{\rho}_k,\hat{\mathbf{J}}_{\perp,k}]\}\geq N_k/|\langle\hat{J}_{z,k}\rangle_{\hat{\rho}_k}|$. Hence, $\xi_{\min}^{2}[\hat{\rho}_{k},\hat{\mathbf{J}}_{\perp,k}]=\lambda_{-,k}<1$ implies that $1>\lambda_{-,k}\geq N_k/|\langle\hat{J}_{z,k}\rangle_{\hat{\rho}_k}|-\lambda_{+,k}\geq 2-\lambda_{+,k}$, whence $\lambda_{+,k}>1$, and we used that $|\langle\hat{J}_{z,k}\rangle_{\hat{\rho}_k}|\leq N_k/2$. 

Assuming that local squeezing is present in each mode, we conclude that it is optimal to encode all $M$ parameters into the respective squeezed local variables that correspond to the eigenvalues $\lambda_{-,k}$ for $k=1,\dots,M$. Formally this is achieved by a transformation matrix of the form
\begin{align}\label{eq:Roptlcl}
\matR=
\begin{pmatrix}r_{x,1} & r_{y,1} & 0 & 0 & \cdots & 0\\
0 & 0 & r_{x,2} & r_{y,2} & \dots & 0\\
\vdots & & &\ddots & &\vdots\\
0 & 0 && \cdots & r_{x,M} & r_{y,M}
\end{pmatrix},
\end{align}
where $\mathbf{r}_k=(r_{x,k},r_{y,k})^T$ is normalized. Minimizing over the local directions $\mathbf{r}_k$ yields with $\hat{\mathbf{H}}=\matR\hat{\mathbf{J}}_{\perp}=\hat{\mathbf{J}}_{\mathbf{r}}$:
\begin{align}\label{eq:locsqzmatoptimizedall}
\xi_{\min}^{2}[\hat{\rho}_{\rm loc},\hat{\mathbf{J}}_{\perp}]:=&\min_{\mathbf{r}_1,\dots,\mathbf{r}_M}\matxi_{\rm opt}^{2}[\hat{\rho}_{\rm loc},\hat{\mathbf{J}}_{\mathbf{r}},\hat{\mathbf{J}}_{\perp}]\\=&\min_{\mathbf{r}_1,\dots,\mathbf{r}_M}\bigoplus_{k=1}^MN_k\mathbf{r}_k^T\tilde{\matM}[\hat{\rho}_k,\hat{\mathbf{J}}_{\perp,k}]^{-1}\mathbf{r}_k\notag\\
=&\bigoplus_{k=1}^M\xi_{\min}^{2}[\hat{\rho}_{k},\hat{\mathbf{J}}_{\perp,k}].\notag
\end{align}
Using Eq.~(\ref{eq:mblocks}), we further find
\begin{align}\label{eq:locsqzoptk}
\xi_{\min}^{2}[\hat{\rho}_{k},\hat{\mathbf{J}}_{\perp,k}]
&=\frac{N_k}{\langle\hat{J}_{z,k}\rangle_{\hat{\rho}_k}^{2}}\left(\Delta_+ - \sqrt{\mathrm{Cov}(\hat{J}_{x,k},\hat{J}_{y,k})_{\hat{\rho}_k}^2 + \Delta_-^2}\right)
\end{align}
with $2\Delta_{\pm}=(\Delta \hat{J}_{x,k})^2_{\hat{\rho}_k} \pm (\Delta \hat{J}_{y,k})^2_{\hat{\rho}_k}$.

Inserting Eq.~(\ref{eq:Roptlcl}) into~(\ref{eq:optimalXS}), it follows immediately that the block-diagonal structures of the matrices $\matGamma[\hat{\rho}_{\rm loc},\hat{\mathbf{J}}_{\perp}]$ and $\tilde{\matC}[\hat{\rho}_{\rm loc},\hat{\mathbf{J}}_{\perp}]$ (which is block-diagonal for all states) allow for a local set of optimal measurement observables. It follows that local parameter encodings and local measurements are indeed optimal for products of locally squeezed states.

We also remark that we may use the form~(\ref{eq:ClclJ}) to show explicitly that for local evolutions $\hat{H}_{\mathbf{r}_k,k}$ and measurements $\hat{X}_{\mathbf{s}_k,k}$, it is favorable to choose the vectors $\mathbf{r}_k$ and $\mathbf{s}_k$ orthogonal to each other and to the mean spin direction $\mathbf{n}_{0,k}$. To see this, note that mode-local $M\times 3M$ transformation matrices $\matR=\mathrm{diag}(\mathbf{r}_1^T,\dots,\mathbf{r}_M^T)$ and $\matS=\mathrm{diag}(\mathbf{s}_1^T,\dots,\mathbf{s}_M^T)$ lead to $\matS\tilde{\matC}[\hat{\rho},\hat{\mathbf{J}}]\matR^T=\mathrm{diag}(\mathbf{s}_1^T(\mathbf{r}_1\times \langle \hat{\mathbf{J}}_1\rangle_{\hat{\rho}}),\dots,\mathbf{s}_M^T(\mathbf{r}_M\times \langle \hat{\mathbf{J}}_M\rangle_{\hat{\rho}}))$. The diagonal elements are therefore proportional to the volume spanned by the unit vectors $\mathbf{r}_k$, $\mathbf{s}_k$ and $\mathbf{n}_{0,k}$, which is maximized by an orthogonal configuration.

\subsubsection{Local spin squeezing matrix}
In summary, in absence of mode correlations, a set of local measurement observables $\hat{\mathbf{J}}_{\mathbf{s}}=(\hat{J}_{\mathbf{s}_1,1},\dots,\hat{J}_{\mathbf{s}_M,M})^T$ and phase-imprinting Hamiltonians $\hat{\mathbf{J}}_{\mathbf{r}}=(\hat{J}_{\mathbf{r}_1,1},\dots,\hat{J}_{\mathbf{r}_M,M})^T$ is optimal. For any choice of $\hat{\mathbf{J}}_{\mathbf{r}}$ and $\hat{\mathbf{J}}_{\mathbf{s}}$, we obtain a diagonal squeezing matrix
\begin{align}\label{eq:SQZloc}
\matxi^{2}[\hat{\rho}_{\rm loc},\hat{\mathbf{J}}_{\mathbf{r}},\hat{\mathbf{J}}_{\mathbf{s}}]=\begin{pmatrix}\xi^{2}[\hat{\rho}_1,\hat{J}_{\mathbf{r}_1,1},\hat{J}_{\mathbf{s}_1,1}] & \cdots & 0\\
\vdots & \ddots  & \vdots & \\
0 & \cdots & \xi^{2}[\hat{\rho}_M,\hat{J}_{\mathbf{r}_M,M},\hat{J}_{\mathbf{s}_M,M}] \end{pmatrix},
\end{align}
and an additional optimization of each of the $\xi^{2}[\hat{\rho}_k,\hat{J}_{\mathbf{r}_k,k},\hat{J}_{\mathbf{s}_k,k}]=N_k(\Delta \hat{J}_{\mathbf{s}_k,k})^{2}_{\hat{\rho}_k}/\langle\hat{J}_{z,k}\rangle_{\hat{\rho}_k}^2$ through the choice of the $\mathbf{r}_k$ and $\mathbf{s}_k$ \cite{MaPHYSREP2011} yields the smallest possible squeezing matrix for this class of states, which is given analytically in Eq.~(\ref{eq:locsqzmatoptimizedall}). 

The condition $\matxi^{2}[\hat{\rho}_{\rm loc},\hat{\mathbf{J}}_{\mathbf{r}},\hat{\mathbf{J}}_{\mathbf{s}}]\geq \matID_M$ is violated already if a single mode is squeezed. If all the local states $\hat{\rho}_k$ are squeezed and therefore satisfy $\xi^{2}[\hat{\rho}_k,\hat{J}_{\mathbf{r}_k,k},\hat{J}_{\mathbf{s}_k,k}]<1$, the full multimode squeezing condition $\matxi^{2}[\hat{\rho}_{\mathrm{loc}},\hat{\mathbf{J}}_{\mathbf{r}},\hat{\mathbf{J}}_{\mathbf{s}}]<\matID$ is met. Such states thus lead to multiparameter sub-shot-noise sensitivities for parameters encoded locally by $\hat{\mathbf{J}}_{\mathbf{r}}$.

\subsection{Nonlocal spin squeezing}
\subsubsection{Local parameter encodings}
We now consider an arbitrary state $\hat{\rho}$ that may contain mode entanglement and analyze the spin squeezing matrix for local parameter encoding schemes that we found to be optimal for local squeezing. As was shown in the main manuscript, in general, the spin squeezing matrix is no longer diagonal and its elements are described by
\begin{align}\label{eq:winelandmatrixS}
(\matxi^2[\hat{\rho},\hat{\mathbf{J}}_{\mathbf{r}},\hat{\mathbf{J}}_{\mathbf{s}}])_{kl}=\frac{\sqrt{N_kN_l}\mathrm{Cov}(\hat{J}_{\mathbf{s}_k,k},\hat{J}_{\mathbf{s}_l,l})_{\hat{\rho}}}{\langle\hat{J}_{z,k}\rangle_{\hat{\rho}}\langle\hat{J}_{z,l}\rangle_{\hat{\rho}}}.
\end{align}
We can now understand under which conditions mode correlations further enhance the sensitivity beyond Eq.~(\ref{eq:SQZloc}). We consider a linear combination of parameters, defined by the coefficients $\mathbf{n}=(n_1,\dots,n_M)^T$. According to Eqs.~(\ref{eq:gaussianfishermatrixS}) and~(\ref{eq:momentmatrixS}), the variance of the estimation of $\mathbf{n}^T\vectheta=\sum_{i=1}^Mn_i\theta_{i}$ is given for an arbitrary quantum state $\hat{\rho}$ by
\begin{align}
\mu\mathbf{n}^T\matSigma\mathbf{n}&=\sum_{k,l=1}^Mn_kn_l\frac{\mathrm{Cov}(\hat{J}_{\mathbf{s}_k,k},\hat{J}_{\mathbf{s}_l,l})_{\hat{\rho}}}{\langle\hat{J}_{z,k}\rangle_{\hat{\rho}}\langle\hat{J}_{z,l}\rangle_{\hat{\rho}}}=\sum_{k,l=1}^Mn_kn_l\frac{(\matxi^2[\hat{\rho},\hat{\mathbf{J}}_{\mathbf{r}},\hat{\mathbf{J}}_{\mathbf{s}}])_{kl}}{\sqrt{N_kN_l}},
\end{align}
and in the second step, we used the definition of the spin squeezing matrix. This sum contains the weighted average of local spin-squeezing coefficients ($k=l$), in addition to the nonlocal squeezing described by covariances ($k\neq l$). It is clear that if the signs of the nonlocal squeezing terms are chosen properly and in accordance with the $n_k$ \cite{GessnerPRL2018}, they can further enhance the sensitivity, as is illustrated by the example in the main text.

\subsubsection{Nonlocal parameter encodings}
In the main text we limited the analysis to local parameter-encoding Hamiltonians in the two spatial modes. For the locally squeezed states of the type~(\ref{eq:SQZloc}), such local phase shifts and measurements are in fact optimal (recall section~\ref{app:lclsqz}): The block diagonal structure of the matrix $\tilde{\matM}[|\Psi_{\rm loc}(t)\rangle,\hat{\mathbf{J}}_{\perp}]$ ensures that the highest multiparameter sensitivity is achieved by encoding each parameter into the respective squeezed local variable via $\hat{\mathbf{H}}=\hat{\mathbf{J}}_{\mathbf{r}}$ and a collection of local observables $\hat{\mathbf{X}}_{\rm opt}=\hat{\mathbf{J}}_{\mathbf{s}}$ saturates the upper bound in Eq.~(\ref{eq:momentsboundapp}). However, as we will see below, such local schemes are generally not optimal in the presence of mode entanglement.

\begin{figure}[tb]
\centering
\includegraphics[width=.49\textwidth]{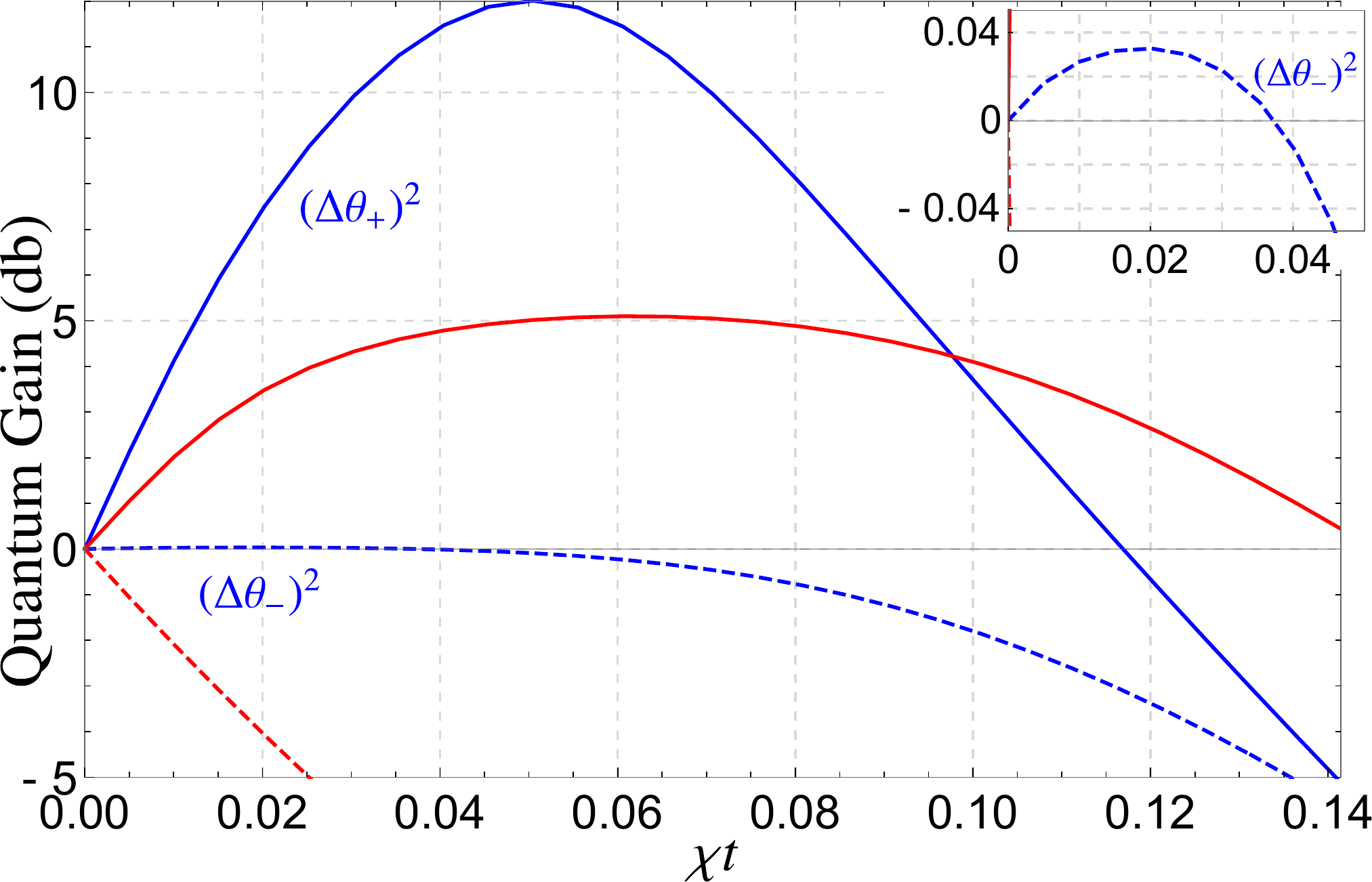}
\caption{Comparison between local and nonlocal squeezing with nonlocal parameter encoding. The plot shows on dB scale $10\log_{10}[(\Delta\theta_+)^2_{\rm SN}/(\Delta\theta_+)^2]$ (continuous lines) and $10\log_{10}[(\Delta\theta_-)^2_{\rm SN}/(\Delta\theta_-)^2]$ (dashed lines) for the nonlocally squeezed state $|\Psi_{\rm nl}(t)\rangle$ (blue lines) with $N=100$ particles. The orientation $\mathbf{r}_1$ and $\mathbf{r}_2$ of the encoding generators have been locally optimized to provide the largest quantum gain. While the quantum gain for $\theta_+$ is maximal, we observe a small gain also for $\theta_-$ at small times (see inset). The locally squeezed state $|\Psi_{\rm loc}(t)\rangle$ provides a weaker quantum gain at relevant short time scales for $\theta_+$, and is above shot noise for $\theta_-$, given the same choice of nonlocal Hamiltonians (red lines).}
\label{fig:NLSQZNLH}
\end{figure}

We now extend our analysis of the nonlocal squeezed state $|\Psi_{\rm nl}(t)\rangle$ to nonlocal measurements  and parameter-imprinting evolutions. Specifically, consider $\hat{\mathbf{H}}=(\hat{H}_1,\hat{H}_2)^T$ with $\hat{H}_1=\frac{1}{\sqrt{2}}(\hat{J}_{\mathbf{r}_1,1}+\hat{J}_{\mathbf{r}_1,2})$ and $\hat{H}_2=\frac{1}{\sqrt{2}}(\hat{J}_{\mathbf{r}_2,1}-\hat{J}_{\mathbf{r}_2,2})$, where $\mathbf{r}_1,\mathbf{r}_2$ are two orthonormal vectors in the $xy$ plane, as well as the measurement observables $\hat{\mathbf{X}}=(\hat{X}_1,\hat{X}_2)^T$ with $\hat{X}_1=\frac{1}{\sqrt{2}}(\hat{J}_{\mathbf{r}_2,1}+\hat{J}_{\mathbf{r}_2,2})$ and $\hat{X}_2=\frac{1}{\sqrt{2}}(\hat{J}_{\mathbf{r}_1,2}-\hat{J}_{\mathbf{r}_1,1})$. 
This scheme describes a nonlocal encoding of two parameters $\vectheta=(\theta_+,\theta_-)^T$ by the evolution $\hat{U}(\vectheta)=\exp(-i\hat{H}_1\theta_+-i\hat{H}_2\theta_-)$. We obtain the squeezing matrix
\begin{align}\label{eq:optnlsqz}
\matxi^{2}[|\Psi_{\rm nl}(t)\rangle,\hat{\mathbf{H}},\hat{\mathbf{X}}]&=\begin{pmatrix}\xi_+^{2} & 0 \\
0 & \xi_-^{2} \end{pmatrix}.
\end{align}
After an optimization over $\mathbf{r}_1$ and $\mathbf{r}_2$, the eigenvalue $\xi_+^2$ coincides with the single-parameter spin-squeezing coefficient for the full ensemble of $N$ atoms: $\xi_+^{2}=\min_{\mathbf{r}_1,\mathbf{r}_2}N(\Delta\hat{J}_{\mathbf{r}_2})^{2}_{|\Psi_{\rm nl}(t)\rangle}/|\langle [\hat{J}_{\mathbf{r}_1},\hat{J}_{\mathbf{r}_2}]\rangle_{|\Psi_{\rm nl}(t)\rangle}|^2$ with $\hat{J}_{\mathbf{r}}=\hat{J}_{\mathbf{r},1}+\hat{J}_{\mathbf{r},2}$, and indicates significant quantum enhancements (blue line in Fig.~\ref{fig:NLSQZNLH}). The quantum gain expressed by it can be achieved through the collective evolution and measurement operators $\hat{H}_1$ and $\hat{X}_1$. In contrast, the gain $\xi^{2}_-$ is accessible only by local measurements on the two spin ensembles. We notice that sub-shot-noise measurements of one parameter do not necessarily imply a reduced sensitivity below the classical limit for the other: At short times, the sensitivity of $\theta_-$ remains close to the shot-noise limit and even slightly undercuts it for very small $\chi t$.  For comparison, the weaker sensitivity of the locally squeezed state $|\Psi_{\rm{loc}}(t)\rangle$ is shown (red lines).

For the nonlocally squeezed state, the choice of phase-imprinting generators $\hat{\mathbf{H}}$ defined above is optimal: The eigenvalues of $\matM_{\mathrm{opt}}[|\Psi_{\mathrm{nl}}(t)\rangle,\mathbf{H},\hat{\mathbf{J}}_{\perp}]$ are maximized by this choice of $\hat{\mathbf{H}}$. Since the shot-noise matrix $\matF_{\rm SN}[\hat{\mathbf{H}}]$ is diagonal, the same $\hat{\mathbf{H}}$ minimize also the eigenvalues of the squeezing matrix $\matxi^2_{\mathrm{opt}}[|\Psi_{\rm nl}(t)\rangle,\hat{\mathbf{H}},\hat{\mathbf{J}}_{\perp}]$. Furthermore, the $\hat{\mathbf{X}}$ satisfy the optimality condition Eq.~(\ref{eq:optimalXS}), leading to $\matxi^{2}[|\Psi_{\rm nl}(t)\rangle,\hat{\mathbf{H}},\hat{\mathbf{X}}]=\matxi_{\rm opt}^{2}[|\Psi_{\rm nl}(t)\rangle,\hat{\mathbf{H}},\hat{\mathbf{J}}_{\perp}]$, where the optimal squeezing matrix was defined in Eq.~(\ref{eq:xioptS}). Moreover, additional $M\times M$ orthogonal transformations $\hat{\mathbf{H}}'=\matV\hat{\mathbf{H}}$ can be used to modify the parameter-encoding evolution with the effect of changing the eigenvectors of $\matxi_{\rm opt}^{2}[|\Psi_{\rm nl}(t)\rangle,\hat{\mathbf{H}},\hat{\mathbf{J}}_{\perp}]$ without changing the eigenvalues. We recall that such transformations have no impact on the optimality of the measurement observables, due to the freedom provided by the matrix $\matT$ in Eq.~(\ref{eq:optimalXS}).

\section*{Supplementary Note 4: Continuous-variable squeezing matrix}\label{app:cvoptsqz}
\setcounter{subsection}{0}
\subsection{Equivalence to the squeezing condition by Simon \textit{et al.}}\label{app:SimonSQZ}
We show that in the context of continuous-variable systems and considering quadrature observables $\hat{\mathbf{q}}$, the squeezing condition, i.e., any violation of
\begin{align}\label{eq:SQZcondS}
\matxi^{2}[\hat{\rho},\hat{\mathbf{H}},\hat{\mathbf{X}}]\geq \matID_M,
\end{align}
becomes equivalent to
\begin{align}\label{eq:sqzSimonS}
\lambda_{\min}(\matGamma[\hat{\rho},\hat{\mathbf{q}}])<\frac{1}{4},
\end{align}
which was proposed by Simon \textit{et al.} in Ref.~\cite{SimonPRA1994}. 

Assume that~(\ref{eq:SQZcondS}) is violated by the matrix $\matxi^{2}[\hat{\rho},\hat{\mathbf{H}},\hat{\mathbf{X}}]$ for some $\hat{\mathbf{H}}=\matR\hat{\mathbf{q}}$ and $\hat{\mathbf{X}}=\matS\hat{\mathbf{q}}$. Since $\matxi_{\mathrm{opt}}^{2}[\hat{\rho},\hat{\mathbf{H}},\hat{\mathbf{q}}]\leq \matxi^{2}[\hat{\rho},\hat{\mathbf{H}},\hat{\mathbf{X}}]$ for all $\hat{\mathbf{X}}$, this implies that also $\matxi_{\mathrm{opt}}^{2}[\hat{\rho},\hat{\mathbf{H}},\hat{\mathbf{q}}]$ violates condition~(\ref{eq:SQZcondS}). This is equivalent to $\lambda_{\min}(\matxi_{\mathrm{opt}}^{2}[\hat{\rho},\hat{\mathbf{H}},\hat{\mathbf{q}}])<1$. Because $\matR$ is an orthogonal projection onto an $M$-dimensional subspace, the matrix $\matxi_{\mathrm{opt}}^{2}[\hat{\rho},\hat{\mathbf{H}},\hat{\mathbf{q}}]$ is a compression of the $2M\times 2M$ matrix $4\matOmega^T\matGamma[\hat{\rho},\hat{\mathbf{q}}]\matOmega$. By the inclusion principle \cite{Bernstein}, we obtain $1>\lambda_{\min}(\matxi_{\mathrm{opt}}^{2}[\hat{\rho},\hat{\mathbf{H}},\hat{\mathbf{q}}])\geq \lambda_{\min}(4\matOmega^T\matGamma[\hat{\rho},\hat{\mathbf{q}}]\matOmega)=4\lambda_{\min}(\matGamma[\hat{\rho},\hat{\mathbf{q}}])$, and we used that $\matOmega$ is an orthogonal matrix. Hence, $\lambda_{\min}(\matGamma[\hat{\rho},\hat{\mathbf{q}}])<1/4$. 

Conversely, assume that $\lambda_{\min}(\matGamma[\hat{\rho},\hat{\mathbf{q}}])<1/4$ holds, then there exists an $\matR$, such that $\lambda_{\min}(\matxi_{\mathrm{opt}}^{2}[\hat{\rho},\hat{\mathbf{H}},\hat{\mathbf{q}}])=\lambda_{\min}(4\matOmega^T\matGamma[\hat{\rho},\hat{\mathbf{q}}]\matOmega)=4\lambda_{\min}(\matGamma[\hat{\rho},\hat{\mathbf{q}}])$. This yields $\lambda_{\min}(\matxi_{\mathrm{opt}}^{2}[\hat{\rho},\hat{\mathbf{H}},\hat{\mathbf{q}}])<1$ and hence the squeezing condition~(\ref{eq:SQZcondS}) can be violated for some $\hat{\mathbf{X}}$.

\subsection{Minimizing the squeezing matrix}\label{app:minsqzevals}
Let $\lambda'_1\leq \dots \leq\lambda'_M$ and $\lambda_1\leq \dots\leq\lambda_{2M}$ 
denote the eigenvalues of $\matxi_{\mathrm{opt}}^{2}[\hat{\rho},\hat{\mathbf{H}}]$ and $4\matOmega^T\matGamma[\hat{\rho},\hat{\mathbf{q}}]\matOmega$, respectively. The inclusion principle yields $\lambda'_k\leq\lambda_k$ for all $k=1,\dots,M$. The minimum spectrum is reached when $\lambda'_k=\lambda_k$ holds for all $k=1,\dots,M$. To achieve this, we choose $\hat{\mathbf{H}}_{\rm opt}=\matR_{\rm opt}\hat{\mathbf{q}}$, by picking the rows of $\matR_{\rm opt}$ as $\mathbf{r}_k=\bs{\lambda}_k$ for $k=1,\dots,M$, where the $\bs{\lambda}_k$ are the eigenvectors of $4\matOmega^T\matGamma[\hat{\rho},\hat{\mathbf{q}}]\matOmega$ with eigenvalue $\lambda_k$. 

For the squeezed vacuum state, the eigenvectors $\bs{\lambda}_k$ form the symplectic orthogonal matrix $\matO\matOmega$, where the columns are ordered in pairs acting on the same mode. The projector $\matP_M$ then selects only the squeezed quadratures from each mode, thereby realizing $\matR_{\rm opt}=\matP_M\matO\matOmega$ as described above. From Eq.~(\ref{eq:optimalXS}) we find the optimal measurement operators as $\hat{\mathbf{X}}_{\rm opt}=\matT\matP_M\matO\matOmega\matOmega^T\matGamma[|\Psi_0\rangle,\hat{\mathbf{q}}]^{-1}\hat{\mathbf{q}}=\matT\matP_M\matO\matGamma[|\Psi_0\rangle,\hat{\mathbf{q}}]^{-1}\matO^T\matO\hat{\mathbf{q}}=4\matT\matP_M\bigoplus_{k=1}^M\mathrm{diag}(e^{2r_k},e^{-2r_k})\matO\hat{\mathbf{q}}$, and we obtain $\hat{\mathbf{X}}_{\mathrm{opt}}=\matP_M\matO\hat{\mathbf{q}}$ by choosing $\matT=\frac{1}{4}\mathrm{diag}(e^{2r_1},\dots,e^{2r_M})$. This choice leads to $\matC[|\Psi_0\rangle,\hat{\mathbf{H}},\hat{\mathbf{X}}]=\frac{1}{2}\matS\matOmega\matR^T=\frac{1}{2}\matP_M\matO\matOmega\matOmega^T\matO^T\matP_M^T=\frac{1}{2}\matID_M$.

\subsection{Changing the basis of the squeezing matrix by passive transformations}\label{app:basischange}
Let us consider a fixed family of encoding Hamiltonians with $\hat{\mathbf{H}}=\matP_M\matO\matOmega\hat{\mathbf{q}}$. Any squeezed vacuum state can be expressed as $\hat{U}_{\matV}|\Psi_{0}\rangle$. We obtain $\matGamma[\hat{U}_{\matV}|\Psi_{0}\rangle,\hat{\mathbf{q}}]=\matV^T\matGamma[|\Psi_{0}\rangle,\hat{\mathbf{q}}]\matV$, where $\matV$ is the symplectic orthogonal matrix that describes the passive transformation $\hat{U}_{\matV}$. For an optimal choice of measurement operators $\hat{\mathbf{X}}$, the state $\hat{U}_{\matV}|\Psi_{0}\rangle$ leads to the squeezing matrix $\matxi^2_{\mathrm{opt}}[\hat{U}_{\matV}|\Psi_{0}\rangle,\hat{\mathbf{H}}]=4\matP_M\matO\matV^T\matGamma[|\Psi_{0}\rangle,\hat{\mathbf{q}}]\matV\matO^T\matP_M^T$. Let $\matY$ be the orthogonal symplectic matrix that yields $4\matGamma[|\Psi_{0}\rangle,\hat{\mathbf{q}}]=\matY^T\bigoplus_{k=1}^M\mathrm{diag}(e^{2r_k},e^{-2r_k})\matY$. Choosing $\matV=\matY^T\matW\matO$, we obtain $\matxi_{\mathrm{opt}}^2[\hat{U}_{\matV}|\Psi_{0}\rangle,\hat{\mathbf{H}}]=\matP_M\matW^T\bigoplus_{k=1}^M\mathrm{diag}(e^{2r_k},e^{-2r_k})\matW\matP_M^T$. The symplectic orthogonal matrix $\matW$ can now be chosen such that
\begin{align}\label{eq:basistransformsqzmat}
\matxi^2_{\mathrm{opt}}[\hat{U}_{\matV}|\Psi_{0}\rangle,\hat{\mathbf{H}}]=\sum_{k=1}^Me^{-2r_k}\mathbf{n}_k\mathbf{n}_k^T,
\end{align}
where $\{\mathbf{n}_k\}_{k=1}^M$ is an arbitrary basis of $\mathbb{R}^M$.

For clarity, let us explicitly construct the matrix $\matW$ that achieves this. We represent the projector as $\matP_M=\sum_{i=1}^M\mathbf{e}_i\mathbf{f}_{2i}^T$, where $\{\mathbf{e}_i\}_{i=1}^M$ and $\{\mathbf{f}_i\}_{i=1}^{2M}$ represent canonical bases of $\mathbb{R}^M$ and $\mathbb{R}^{2M}$, respectively. We define $\mathbf{m}_{2i}=(0,n_{i1},\dots,0,n_{iM})^T$ and $\mathbf{m}_{2i-1}=(n_{i1},0,\dots,n_{iM},0)^T$ for $i=1,\dots,M$. The $\{\mathbf{m}_i\}_{i=1}^{2M}$ form a basis of $\mathbb{R}^{2M}$. 
By choosing $\matW=\sum_{i=1}^{2M}\mathbf{m}_i\mathbf{f}_i^T$, we obtain the squeezing matrix 
provided in Eq.~(\ref{eq:basistransformsqzmat}). By construction the matrix $\matW$ is orthogonal. By writing $\matOmega=\sum_{i=1}^M(\mathbf{f}_{2i-1}\mathbf{f}_{2i}^T-\mathbf{f}_{2i}\mathbf{f}_{2i-1}^T)$ and making use of $\sum_{i=1}^Mn_{ki}n_{li}=\mathbf{n}_k^T\mathbf{n}_l=\delta_{kl}$, it is possible to demonstrate explicitly that $\matW\matOmega\matW^T=\matOmega$ and thus $\matW$ is symplectic.
Finally, let us consider the optimal measurement observables $\hat{\mathbf{X}}=\matT\matP_M\matO\matGamma[\hat{U}_{\matV}|\Psi_0\rangle,\hat{\mathbf{q}}]^{-1}\hat{\mathbf{q}}=4\matT\matP_M\matW^T\bigoplus_{k=1}^M\mathrm{diag}(e^{-2r_k},e^{2r_k})\matW\matO\hat{\mathbf{q}}$. Choosing $\matT=\frac{1}{4}\mathrm{diag}(e^{2r_1},\dots,e^{2r_M})$ yields $\hat{\mathbf{X}}=\matP_M\matO\hat{\mathbf{q}}$. It is interesting to notice that the optimal measurement observables are thus independent of the basis that is chosen by $\matW$. Instead they depend only on $\matO$, i.e., the phase-imprinting generators $\hat{\mathbf{H}}$.

\subsection{Optimality of squeezed vacuum states}
In Ref.~\cite{Lang} the intuition about the optimality of squeezed vacuum states for single-parameter estimation with continuous-variable systems~\cite{JooPRL2011} was confirmed by a rigorous demonstration. We now show that analogous results hold for multiparameter estimation problems, where for the optimization we distinguish between the two cases discussed in the previous section.

Let us first discuss the optimization of the spectrum of the quantum Fisher matrix $\matF_Q[\hat{\rho},\hat{\mathbf{H}}]$. For mode-separable probe states the upper sensitivity limit is given by the block-diagonal covariance matrix $\max_{\hat{\rho}_{\rm m-sep}}\matF_Q[\hat{\rho}_{\rm m-sep},\hat{\mathbf{H}}]=4\bigoplus_{k=1}^M\max_{|\psi_k\rangle}(\Delta \hat{H}_k)^2_{|\psi_k\rangle}$ and the convexity of $\matF_Q$ allows us to limit the optimization to pure states \cite{GessnerPRL2018}. Since each $\hat{H}_k$ is a local quadrature operator, the local variances satisfy $4(\Delta \hat{H}_k)^2_{|\Psi_k\rangle}\leq 2N_k+1+2|\langle \hat{a}_k\hat{a}_k\rangle_{|\psi_k\rangle}|$ with $N_k=\langle \hat{a}^{\dagger}\hat{a}\rangle_{|\psi_k\rangle}$. This bound can be derived by taking the larger of the two eigenvalues of the covariance matrix $\matGamma[|\psi_k\rangle,(\hat{x}_k,\hat{p}_k)^T]$, and by setting all mean values to zero (which can only increase the covariance matrix). From the Cauchy-Schwarz inequality we obtain 
$|\langle \hat{a}_k\hat{a}_k\rangle_{|\psi_k\rangle}|^2\leq \langle \hat{a}_k^{\dagger}\hat{a}_k\rangle_{|\psi_k\rangle}\langle \hat{a}_k\hat{a}_k^{\dagger}\rangle_{|\psi_k\rangle}=N_k(N_k+1)$. This finally yields $\max_{|\psi_k\rangle}4(\Delta \hat{H}_k)^2_{|\psi_k\rangle}=2N_k+1+2\sqrt{N_k(N_k+1)}$. This upper limit is saturated by a non-displaced squeezed vacuum state, as can be easily verified using $2N_k+1=\cosh 2r_k$ and $\sinh 2r_k = \pm 2\sqrt{N_k(N_k+1)}$. 

Next, we consider the estimation of a specific linear combination of parameters, defined by the coefficient vector $\mathbf{n}\in\mathbb{R}^M$. The variance $\mu\mathbf{n}^T\matSigma\mathbf{n}=(\mathbf{n}^T\matF_Q[\hat{\rho},\hat{\mathbf{H}}]\mathbf{n})^{-1}$ is minimized by a pure state with $\matF_Q[|\Psi\rangle,\hat{\mathbf{H}}]=4\matGamma[|\Psi\rangle,\hat{\mathbf{H}}]$. Assuming $\mathbf{n}$ to be normalized to one, the sensitivity limit is given by the largest eigenvalue of the $M\times M$ covariance matrix $\max_{|\Psi\rangle}4\matGamma[|\Psi\rangle,\hat{\mathbf{H}}]$, where $\hat{\mathbf{H}}=\matP_M\matO\hat{\mathbf{q}}$. It is achieved when $\mathbf{n}$ represents the corresponding eigenvector. Since $\matO$ is a canonical transformation, each eigenvalue of $\matGamma[|\Psi\rangle,\hat{\mathbf{H}}]$ corresponds to the variance of some quadrature observable that is constructed as a linear combination of the original $\hat{\mathbf{q}}$ and follows the same commutation relations. Following the arguments from above, we obtain the bound $\lambda_{\max}(4\matGamma[|\Psi\rangle,\hat{\mathbf{H}}])\leq 2N+1+2\sqrt{N(N+1)}$, which is again saturated by squeezed vacuum states. Here $N=\sum_{k=1}^MN_k$ is the total number of particles. 

This extends the results of Ref.~\cite{Lang} and demonstrates the optimality of squeezed vacuum also in the multiparameter case. Similar strategies are optimal also for more general multimode passive Gaussian channels that encode a single~\cite{MatsubaraNJP2019} or multiple phases~\cite{OhArxiv}. If transformations beyond displacements are considered, however, the preparation of the optimal state is no longer independent of the values of the unknown phases and thus requires adaptive methods.

\end{document}